\documentclass[12pt]{article}

\usepackage[T1]{fontenc}

\usepackage{amsmath}
\usepackage{amssymb}
\usepackage{slashed}
\usepackage{tikz}
\usepackage[numbers,sort&compress]{natbib}
\usepackage{hyperref}
\usepackage{cleveref}
\crefname{equation}{Eq.}{Eqs.}
\crefname{figure}{Fig.}{Figs.}
\crefname{table}{Table}{Tables}
\crefname{section}{Section}{Sections}

\usepackage[letterpaper,margin=1in,bottom=1in]{geometry}
\usepackage{float} 
\usepackage{parskip} 
\usepackage{tabulary} 
\usepackage{color} 
\usepackage{soul} 
\usepackage{subfig}
\usepackage{graphicx}
\usepackage[section]{placeins} 
\usepackage{multirow}

\def\lhc2{LHC~Run~II}

\newcommand{\code}[1]{\texttt{#1}}

\newcommand{\ifb}{~\textrm{fb}^{-1}}

\def\beq{\begin{equation}}
\def\be{\begin{equation}}
\def\beqn{\begin{eqnarray}}
\def\ee{\end{equation}}
\def\eeq{\end{equation}}
\def\eeqn{\end{eqnarray}}

\begin{document}

\author{Amin Aboubrahim$^b$\footnote{Email: a.abouibrahim@northeastern.edu}~, Wan-Zhe Feng$^a$\footnote{Email: vicf@tju.edu.cn}~and Pran Nath$^b$\footnote{Email: p.nath@northeastern.edu}\\~\\
$^{a}$\textit{\normalsize Center for Joint Quantum Studies and Department of Physics,}\\
\textit{\normalsize School of Science, Tianjin University, Tianjin 300350, PR. China}\\
$^{b}$\textit{\normalsize Department of Physics, Northeastern University,
Boston, MA 02115-5000, USA} \\}

\title{Expanding the parameter space of natural supersymmetry}

\date{}
\maketitle

\begin{abstract}
SUSY/SUGRA models with naturalness defined via small $\mu$ are constrained due to experiment on the relic density
and the experimental limits on the WIMP-proton cross-section and WIMP annihilation cross-section from indirect detection experiments. Specifically models with small $\mu$ where the neutralino is higgsino-like
lead to dark matter relic density below the observed value.
In several works this problem is overcome by assuming dark matter to be constituted of more than one component and the
neutralino relic density deficit is made up from contributions from other components. In this work
we propose that the dark matter consists of just one component, i.e., the lightest neutralino and the relic density of the
higgsino-like neutralino receives contributions from the usual freeze-out mechanism along with contributions
arising from the decay of hidden sector neutralinos.  The model we propose is an extended
MSSM model where the hidden sector is constituted of a $U(1)_X$ gauge sector along with matter charged under
$U(1)_X$ which produce two neutralinos in the hidden sector.
The $U(1)_X$ and the hypercharge $U(1)_Y$  of the MSSM  have kinetic and Stueckelberg mass mixing
where the mixings are ultraweak. In this case the hidden sector neutralinos have ultraweak interactions with
the visible sector. Because of their ultraweak interactions the hidden sector neutralinos are not thermally
produced and we assume their initial relic density to be negligible.
However, they can be produced via interactions of MSSM particles in the early universe, and once produced they decay to the neutralino.
For a range of mixings the decays occur before the BBN producing
additional relic density for the neutralino.
Models of this type are testable in dark matter direct and indirect detection experiments and at the high luminosity and high energy LHC.

\end{abstract}

\section{Introduction}\label{sec:intro}

The observation of the Higgs boson mass at $\sim 125$ GeV~\cite{Aad:2012tfa,Chatrchyan:2012ufa}
implies that the size of weak scale supersymmetry is large lying in the TeV region. The relative size of weak scale supersymmetry implies that the observation of supersymmetry would be more difficult than previously thought. However, the sparticle spectrum
is governed by more than one mass parameter. Specifically if the universal scalar mass lies in the several TeV region,
the sfermion masses are expected to be large. However, the electroweakinos could be much lighter than
the sfermions. This is so because the electroweakino masses for the universal supergravity (SUGRA) model are determined by the
universal gaugino mass at the grand unification scale and the Higgs mixing parameter $\mu$. For the case when
$\mu$ is relatively small the composition of electroweakinos is strongly influenced by the size of $\mu$ and the smaller
$\mu$ is the larger the Higgsino content of electroweakinos.
Of specific interest is the composition of the lightest neutralino
where a small $\mu$ would lead to a higgsino-like neutralino. However, a higgsino-like neutralino has copious annihilation
in the early universe which leads to the relic density of higgsino-like neutralino to fall below the experimental value
of $\Omega h^2\sim 0.12$.
Now small $\mu$ models arise quite naturally on the hyperbolic branch of
radiative breaking of the electroweak symmetry~\cite{Chan:1997bi,Chattopadhyay:2003xi,Akula:2011jx}
(for related works see~\cite{Feng:1999mn,Baer:2003wx,Feldman:2011ud,Ross:2017kjc})
and lead to an electroweakino mass spectrum
which in part lies far below the sfermion masses and would be accessible at future colliders.
A relatively small $\mu$ is also associated with naturalness.
Thus what is natural is to a degree rather subjective
and there are various technical definitions quantifying naturalness.
Typically most naturalness models feature a relatively small $\mu$,
and it is the relative smallness of $\mu$ (compared to, for example, squark masses)
that we will use as the criterion of naturalness in this work.
However, the content of the analysis given in this work stands on its merits
irrespective of the nomenclature one assigns to the models considered.
As indicated above higgsino-like neutralino typically leads to a relic density that falls below the experimental value.
One way to overcome this problem is to assume that  dark matter is made of more than one component~\cite{Feldman:2010wy,Feldman:2011ms,Baer:2018rhs,Aboubrahim:2019mxn}
with each component contributing only a fraction of the total relic density.  In this case the relic density contribution of the
higgsino-like neutralino does not pose a problem as the
deficit can be attributed to other component(s) of dark matter. For instance the other component could be
an axion~\cite{Baer:2018rhs,Halverson:2017deq}
or a Dirac fermion~\cite{Feldman:2010wy,Feldman:2011ms,Aboubrahim:2019mxn}.

In this work we take a different approach. We assume that there is just one component of dark matter and it is the lightest
neutralino which is the lightest supersymmetric particle (LSP). In this case we propose that the relic density arises from two sources: First we have the conventional
freeze-out relic density for the neutralino. Second there is additional contribution to the relic density where
the hidden sector neutralinos decay into the LSP.
The hidden sector neutralinos are assumed to have negligible initial abundance, and
are produced via interactions of the minimal supersymmetric standard model (MSSM) particles in the early universe.
We also assume their masses are larger than the lightest neutralino.
The interactions of the hidden sector neutralinos are ultraweak so they are long-lived but for a range of
ultraweak couplings they decay to the LSP before the Big Bang Nucleosynthesis (BBN) sets in.
In the specfic model we propose, the hidden sector
is constituted of $U(1)_X$ gauge fields and matter fields charged under the $U(1)_X$,
while the hidden sector is not charged under the Standard Model gauge group.
We assume that the interactions between the hidden sector and the
visible sector arise due to kinetic mixing~\cite{Holdom:1985ag,Holdom:1991}
and the Stueckelberg mass mixing~\cite{Kors:Nath,st-mass-mixing,WZFPN,Feldman:2007wj}
between the gauge field of the $U(1)_Y$
hypercharge and the gauge field of the $U(1)_X$.

The outline of the rest of the paper is as follows: In Section~\ref{sec:model} we discuss details of the extended
MSSM/SUGRA model~\cite{msugra,Nath:2016qzm}.
The analysis of the relic density of dark matter in the extended model is given in Section~\ref{sec:DM}.
Here it is shown that the ultraweakly interacting particles produced in the early universe,
i.e., in the post inflationary period, decay into the
LSP of the MSSM and for a range of the parameter space they decay before the BBN time
producing the desired relic density observed today. We consider three classes of processes for the production of
the ultraweakly interacting hidden sector particles which we label as $\xi$~\cite{Hall:2009bx,Belanger:2018mqt,Tsao:2017vtn,Aboubrahim:2019kpb}.
These are: $A+B\to \xi$, $A\to B+\xi$,
$A+B\to C+\xi$ where particles $A,B,C$ are MSSM particles in the thermal bath,
while $\xi$ is a particle not in the thermal bath in the early universe and is assumed to have negligible initial abundance,
and  has a mass larger than the lightest MSSM neutralino.
In Section~\ref{sec:implement} we present the results of the scan performed on the model's parameter space and give a set of benchmarks  which satisfy the Higgs boson mass constraint, the
relic density constraint and are chosen such that the sparticle spectrum
satisfies the current experimental lower limits given by the  LHC.  A discussion on the electroweakino spectrum along with a full collider analysis of the benchmarks are carried out in Section \ref{sec:collider}. A part of the parameter space discussed here can be probed at HL-LHC and HE-LHC~\cite{CidVidal:2018eel,Cepeda:2019klc,Benedikt:2018ofy,Zimmermann:2018koi} (for related works on HL-LHC and HE-LHC, see Refs.~\cite{Aboubrahim:2018bil,Aboubrahim:2018tpf,Aboubrahim:2019qpc,Aboubrahim:2019vjl,Aboubrahim:2019mxn}).
Conclusions are given in Section~\ref{sec:conc}.
In the Appendix we list  $A+B\to C+\xi$ type processes that contribute to the dark matter relic density.

\section{The model}\label{sec:model}

The model we discuss  contains the visible sector, a hidden sector and the interactions  of the visible sector with the hidden sector so that the total Lagrangian of the extended system has the form~\cite{Nath:1996qs}
\begin{align}
\mathcal{L}= \mathcal{L}_{\rm vis} + \mathcal{L}_{\rm hid} + \mathcal{L}_{\rm vh}\,.
\label{2.1}
\end{align}
In our analysis we will assume that the visible sector is constituted of the MSSM Lagrangian. There are many options for the
hidden sector but to be concrete we will assume that the hidden sector consists of a $U(1)_X$ gauge field and matter charged
under the $U(1)_X$ but  the hidden sector is not charged under the Standard Model gauge group. We assume that
$\mathcal{L}_{\rm vh}$ arises from two sources: First there is a  gauge kinetic mixing between $U(1)_X$ of the hidden sector and
the hypercharge $U(1)_Y$ of the Standard Model gauge group, and additionally there is a Stueckelberg mass mixing
between the $U(1)_Y$ gauge field $B_\mu$ and the $U(1)_X$ gauge field $C_\mu$. Thus for $\mathcal{L}_{\rm vh}$ we have
\begin{align}
\mathcal{L}_{\rm{vh}} = &
-\frac{\delta}{2}B^{\mu\nu}C_{\mu\nu}-i\delta(\lambda_{C}\sigma^{\mu}\partial_{\mu}\bar{\lambda}_B
+\lambda_{B}\sigma^{\mu}\partial_{\mu}\bar{\lambda}_{C})
 -\frac{1}{2}(M_1 C_{\mu}+M_2 B_{\mu}+\partial_{\mu}a)^2 \,, \label{maxmix}
\end{align}
where $\delta$ is the kinetic mixing parameter, $\lambda$ is the gaugino component of the vector superfield.
The axion field $a$ is from the two additional chiral superfields $S$ and $\bar S$ that enter  the model~\cite{Kors:Nath}.
In the unitary gauge  the axion $a$ is absorbed to generate mass for the $U(1)_X$ gauge boson. A more detailed discussion of
the Stueckelberg extended model can be found in~\cite{Kors:Nath,st-mass-mixing,WZFPN,Feldman:2007wj}.
In addition to the above we add soft terms to the Lagrangian so that
\begin{equation}
\Delta\mathcal{L}_{\rm soft} \
=-\left(\frac{1}{2}m_X\bar{\lambda}_X\lambda_X+ M_{XY}\bar{\lambda}_X\lambda_Y\right)\,,
\end{equation}
where $m_X$ is mass of the $U(1)_X$ gaugino and $M_{XY}$ is the $U(1)_X$-$U(1)_Y$ gaugino mixing mass.
We note that the mixing parameter $M_{XY}$ and $M_2$ even when set to zero at the grand unification scale will assume
non-vanishing values due  to renormalization group evolution.
Thus $M_{XY}$ has the beta-function evolution so that
\begin{equation}
\beta^{(1)}_{M_{XY}}=\frac{33}{5}g^2_Y\left[M_{XY}-(M_1+m_X)s_{\delta}+M_{XY}s^2_{\delta} \right]\,,
\end{equation}
where $g_Y$ is the $U(1)_Y$ gauge coupling and $s_{\delta}=\delta/(1-\delta^2)^{1/2}$.
Similarly, the  mixing parameter $M_2$ has the beta-function so that
\begin{equation}
\beta^{(1)}_{M_2}=\frac{33}{5}g^2_Y(M_2-M_1 s_{\delta})\,,
\label{m2rge}
\end{equation}
In the MSSM sector we will take the soft terms to consist of $m_0, ~A_0, ~m_1, ~m_2, ~m_3, ~\tan\beta,
 ~\text{sgn} (\mu)$. Here
 $m_0$ is the universal scalar mass, $A_0$ is the universal trilinear coupling, $m_1,  ~m_2,  ~m_3$ are the masses of the $U(1)_Y$, $SU(2)_L$, and $SU(3)_C$ gauginos, $\tan\beta=v_u/v_d$ is the ratio of the Higgs vacuum expectation values and $\text{sgn}(\mu)$ is the sign of the Higgs mixing parameter which is chosen to be positive. Here we have assumed non-universalities in the gaugino mass sector which will be useful in the analysis
 in Section \ref{sec:implement} (for some relevant works on non-universalities in the gaugino masses
 see Ref.~\cite{nonuni-gaugino}).

The neutralino sector of the extended SUGRA model contains 6 neutralinos.
We label the mass eigenstates as
$\tilde\xi^0_1, ~\tilde\xi^0_2; ~\tilde \chi_1^0, ~\tilde \chi_2^0, ~\tilde \chi_3^0, ~\tilde \chi_4^0\,$.
Since the mixing parameter $\delta$ is assumed to be
very small ($\delta\lesssim 10^{-10}$),
the first two neutralinos $\tilde\xi^0_1$ and $\tilde\xi^0_2$ reside mostly in the hidden sector
while the remaining four $\tilde \chi_i^0$ ($i=1\cdots 4$) reside mostly in the MSSM sector.
The details of the mixing are given in \cite{Aboubrahim:2019vjl,Aboubrahim:2019kpb}.
For the classes of models we are interested in,
$\tilde \chi_1^0$ is the LSP of the entire supersymmetric sector and thus the dark matter candidate.
Although the coupling between the MSSM sector and the hidden sector is ultraweak,
MSSM particles can produce significant amount of $\tilde\xi^0_1, \tilde\xi^0_2$,
and once produced $\tilde\xi^0_1, \tilde\xi^0_2$ will subsequently decay to the LSP $\tilde \chi_1^0$.
These decay processes would happen after the lightest neutralino $\tilde\chi^0_1$ goes through the freeze-out process
and thus will make additional contribution to the $\tilde\chi^0_1$ relic density.
The lifetime of the hidden sector neutralinos $\tilde\xi^0_1, \tilde\xi^0_2$ is less than 1-10 second,
and their late decay will  still be consistent with the BBN.
Thus, the relic density of $\tilde \chi_1^0$ dark matter consists of two parts:
  (1) The normal $\tilde \chi_1^0$ freeze-out contribution; (2) A contribution arising from the decays of the hidden sector neutralinos $\tilde\xi^0_1, \tilde\xi^0_2$ to $\tilde \chi_1^0$. As will  be seen in our analysis later, the  contribution (2) is very significant in achieving the desired relic density for dark matter.

We turn now to the charge neutral gauge vector boson sector. Here the $2\times 2$ mass-squared matrix of the Standard Model
is enlarged to become a $3\times 3$ mass-squared matrix in the $U(1)_X$-extended SUGRA model.
Thus  after spontaneous electroweak symmetry breaking and  the Stueckelberg mass growth the
mass-squared matrix of neutral vector bosons is a $3\times 3$  matrix
in the basis $(C_{\mu}, B_{\mu}, A^3_{\mu})$ where $A_{\mu}^3$ is the
neutral component of the $SU(2)_L$ gauge field $A_\mu^a$, $a=1-3$.
This $3\times 3$ matrix has three eigenstates which are the photon, the $Z$ boson and the $Z^\prime$ boson.
Assuming that the hidden sector neutralinos have masses greater than $\tilde\chi^0_1$, they will decay to the
$\tilde\chi^0_1$ via interactions involving the $Z,Z^\prime$ and also via Higgs interactions. Computations of these interactions
are straightforward extensions of the MSSM interactions and details of how this can be carried out  can be
found in~\cite{Kors:Nath,st-mass-mixing,WZFPN,Feldman:2007wj}.

\section{Dark matter relic density}\label{sec:DM}

As noted in the introduction, models with small  $\mu$ such that the lightest neutralino has a significant higgsino content
have the problem of not getting enough relic density for the neutralinos as they annihilate copiously in the early universe.
 One way to overcome this problem is to have multi-component dark matter
where the deficit is made up from other dark matter candidates.  Also as mentioned in the introduction, in this
work we propose another possibility where the ultraweakly interacting particles in the hidden sector
decay into the neutralino to make up the deficit. Thus the relic density in this case consists of two parts so that
\begin{equation}
\Omega h^2 = (\Omega h^2)_{1} + (\Omega h^2)_{2}\,,
\label{totrelic}
\end{equation}
where $(\Omega h^2)_{1}$ is the relic density arising from the usual freeze-out mechanism while $(\Omega h^2)_{2}$
is the relic density arising from the decay of the hidden sector neutralinos into the MSSM neutralino.
 In terms of the comoving number density $Y$, the relic density of a dark matter particle of mass $m$ is given by
\begin{equation}
\Omega h^2=\frac{m Y s_0 h^2}{\rho_c}\,,
\label{reliceq}
\end{equation}
where $s_0$ is today's entropy density, $\rho_c$ is the critical density and $h=0.678$.
Below we discuss the main contributions to $(\Omega h^2)_{2}$.
The contributions to  $(\Omega h^2)_{2}$ arise mostly from $A\to B+\xi$ type processes
where particles $A$ and $B$ are MSSM particles and $\xi$ is a hidden sector particle which
is assumed heavier than the LSP and decays into it before the BBN time.
In a similar fashion we also have $A+B\to \xi$ and $A+B\to C+\xi$ types of processes
where $C$ is also an MSSM particle in the thermal bath.
We assume that all the processes above occur when bath particles $A,B,C$ are
in thermal equilibrium in the early universe.
We discuss these processes in further detail below.

\noindent {\bf{$A\to B+\xi$ process:}}

Here a bath particle $A$ decays to another bath particle $B$ plus the hidden sector particle $\xi$.
In the model we discuss here $\xi$ could be the hidden sector neutralino $\tilde\xi^0_1$ or $\tilde\xi^0_2$
where $A$ and $B$ are in thermal equilibrium, while $\xi$ is not and we assume it  has a negligible initial abundance.
In this case the Boltzmann equation for the number density of $\xi$ is given by
\begin{align}
\dot{n}_{\xi}+3Hn_{\xi} & =\int{\rm d}\Pi_{\xi}{\rm d}\Pi_{A}{\rm d}\Pi_{B}(2\pi)^{4}\delta^{4}(p_{A}-p_{B}-p_{\xi})\nonumber \\
 & \quad\times\left[|M|_{A\to B+\xi}^{2}f_{A}(1\pm f_{B})(1\pm f_{\xi})-|M|_{B+\xi\to A}^{2}f_{B}f_{\xi}(1\pm f_{A})\right]\,,\label{Bol1to2}
\end{align}
where ${\rm d}\Pi_{i}=\frac{{\rm d}^{3}\mathbf{p}_{i}}{(2\pi)^{3}2E_{i}}$
are phase space elements,  $f_{i}=[{\rm exp}(E_{i}-\mu_{i})/T\pm1]^{-1}$ is the phase space density.
The plus sign in the above expression is for 
 bosons and minus for fermions. In Eq.~(\ref{Bol1to2}), the matrix element squared $|M|^{2}$ is summed over initial and final spin and color states.
The initial $\xi$ abundance being zero indicates $f_{\xi}=0$,
and thus the term corresponding to $B+\xi \to A$ in Eq.~(\ref{Bol1to2}) vanishes.
By setting $(1\pm f_{B})\approx1$, Eq.~(\ref{Bol1to2}) reduces to
\begin{equation}
\dot{n}_{\xi}+3Hn_{\xi}  =\int{\rm d}\Pi_{\xi}{\rm d}\Pi_{A}{\rm d}\Pi_{B}(2\pi)^{4}\delta^{4}(p_{A}-p_{B}-p_{\xi})|M|_{A\to B+\xi}^{2}f_{\xi}\,.\label{Bol1to2-1}
\end{equation}
Further one can reduce Eq.~(\ref{Bol1to2-1}) so that it takes the form 
\begin{equation}
\dot{n}_{\xi}+3Hn_{\xi} \approx g_{A}\int\frac{{\rm d}^{3}\mathbf{p}_{A}}{(2\pi)^{3}2E_{A}}2m_{A}\Gamma_{A}f_{A}
\approx 2 g_{A}m_{A}\Gamma_{A}\int\frac{{\rm d}^{3}\mathbf{p}_{A}}{(2\pi)^{3}2E_{A}}e^{-E_{A}/T}\,,\label{Bol1to2-2}
\end{equation}
where $\Gamma_A$ is the $A\to B+\xi$ partial decay width
and we have used $f_{A}\approx e^{-E_{A}/T}$.
Changing the differentiation variable to energy
we can further write
\begin{equation}
\dot{n}_{\xi}+3Hn_{\xi} \approx\frac{g_{A}m_{A}\Gamma_{A}}{2\pi^{2}}\int_{m_{A}}^{\infty}\sqrt{E_A^{2}-m_A^{2}}e^{-E_{A}/T}{\rm d}E_{A}
 =\frac{g_{A}m_{A}^{2}\Gamma_{A}}{2\pi^{2}}TK_{1}\left(\frac{m_{A}}{T}\right)\,,
\end{equation}
where $K_{1}$ is the Bessel function of the second kind and degree one,
which is given by the integral
\begin{equation}
K_{1}(z)=z\int_{1}^{\infty}e^{-zx}\sqrt{x^{2}-1}\,{\rm d}x\,. \label{bessel}
\end{equation}
Again using the conservation of entropy per comoving volume ($sR^{3}={\rm const}$),
we have
\begin{equation}
\dot{n}_{\xi}+3Hn_{\xi}=s\dot{Y}_{\xi}\,,
\end{equation}
where $Y_{\xi}\equiv n_{\xi}/s$ the number density of $\xi$ per comoving
volume. Defining $x_A\equiv m_A/T$ we arrive finally
\begin{align}
Y_{\xi} \approx \frac{g_{A}}{2\pi^2}\Gamma_A m^2_A \int_{T_{\rm min}}^{T_{\rm max}}  \frac{{\rm d}T}{s(T)H(T)} K_1(x_A)\,,
\label{Yx2}
\end{align}
where the entropy density and the Hubble parameter are given by
\begin{align}
s(T) & =\frac{2\pi^{2}}{45}T^{3}g_{*S}\,,  \label{ST}\\
H(T) &\approx1.66\sqrt{g_{*}}\frac{T^{2}}{M_{{\rm pl}}}\,. \label{HT}
\end{align}
In Eqs.~(\ref{ST}) and (\ref{HT}), $g_{*S}$ and $g_{*}$ are the effective number of degrees of freedom at temperature $T$ for the entropy and energy density, respectively, and $M_{\rm pl}$ is the Planck mass.
We introduce the fugacity $z$ of the system as
$z=z_f e^{\mu_c/T}$ with $\mu_c$ being the chemical potential
and $z_f = +1$ for a boson, $-1$ for a fermion and zero for a dark matter particle.
In the numerical analysis we use the more exact form of $Y_{\xi}$ given by
\begin{equation}
Y_{\xi}=\frac{g_{A}|z_A|}{2\pi^2}\Gamma_A m^2_A \int_{T_0}^{T_R}\frac{{\rm d}T}{H'(T)s(T)}K'_1(x_A,x_{\xi},x_B,z_A,z_{\xi},z_B)\,,
\label{yield}
\end{equation}
where $T_0$ is the current temperature and $T_R$ is the reheating temperature and we have defined $K'_1$ as the generalized Bessel function of the second kind of degree one given by
\begin{equation}
K'_1(x_A,x_{\xi},x_B,z_A,z_{\xi},z_B)=x_A\int_1^{\infty}\frac{du\sqrt{u^2-1}e^{-x_A u}}{1-z_A e^{-x_A u}}S(x_A\sqrt{u^2-1},x_A,x_{\xi},x_B,z_{\xi},z_B)\,,
\label{genBessel}
\end{equation}
with the function $S$ is as defined in~\cite{Belanger:2018mqt} and is given by
\begin{equation}
S(p_A/T,x_A,x_{\xi},x_B,z_{\xi},z_Y)=\frac{1+\frac{m_A T}{2p_A p_{\xi,B}}\log\left[\frac{(1-z_B e^{-E_B(1)/T})(1-z_{\xi} e^{-E_{\xi}(-1)/T})}{(1-z_{\xi} e^{-E_{\xi}(1)/T})(1-z_B e^{-E_B(-1)/T})}\right]}{1-z_B z_{\xi}e^{-E_A/T}}\,.
\end{equation}
We note that 
 neglecting the effect of the chemical potential, i.e. setting $z_B$ and $z_{\xi}$ to zero, $S\rightarrow 1$ and  Eq.~(\ref{genBessel}) reduces to Eq.~(\ref{bessel}). The function $K_1'$,  which takes six arguments corresponding to values of
$x_{A,\xi,B}$ where $x=m/T$ and  corresponding to the  fugacity parameters $z_{A,\xi,B}$,  is evaluated using \code{micrOMEGAs5.0} routines. The integral of Eq.~(\ref{yield}) is then computed to determine the relic density of $\xi$ using Eq.~(\ref{reliceq}). The hidden sector particle will eventually decay to the dark matter particle of mass $m_{\rm DM}$ for which the relic density is given by
\begin{equation}
(\Omega h^2)_{\rm DM}=\frac{m_{\rm DM}}{m_{\xi}}\,(\Omega h^2)_{\xi}\,.
\end{equation}

\noindent {\bf{$A+B\to\xi$  process:}}

Here two bath particles in thermal equilibrium combine into the hidden sector particle $\xi$ which could be
$\tilde\xi^0_1$ or $\tilde\xi^0_2$. For example, 
Higgs or $Z$ boson combine with a light neutralino so that $\tilde\chi^0_1 + h/Z \to \tilde\xi^0_1$, or  $W^{\pm}$ boson combine with charginos,
so that  $\tilde\chi^{\pm}_1 + W^{\mp} \to \tilde\xi^0_1$, etc.
The Boltzmann equation for the number density of $\xi$ in this case reads
\begin{align}
\dot{n}_{\xi}+3Hn_{\xi} & =\int{\rm d}\Pi_{\xi}{\rm d}\Pi_{A}{\rm d}\Pi_{B}(2\pi)^{4}\delta^{4}(p_{\xi}-p_{A}-p_{B})\nonumber \\
 & \quad\times\left[|M|_{A+B\to \xi}^{2}f_{A}f_{B}(1\pm f_{\xi})-|M|_{\xi\to A+B}^{2}f_{\xi}(1\pm f_{A})(1\pm f_{B})\right]\,,\label{Bol2to1}
\end{align}
where the second term in the parenthesis can be dropped since the
initial abundance of $\xi$ is negligible.
Using the principle of detailed balance, one can rewrite Eq.~(\ref{Bol2to1}) as 
\begin{equation}
\dot{n}_{\xi}+3Hn_{\xi}
\approx\int{\rm d}\Pi_{\xi}{\rm d}\Pi_{A}{\rm d}\Pi_{B}(2\pi)^{4}\delta^{4}(p_{\xi}-p_{A}-p_{B})
|M|_{\xi\to A+B}^{2}f_{\xi}^{{\rm EQ}}\,, \label{Bol2to1-2}
\end{equation}
where $f_{\xi}^{{\rm EQ}} \approx e^{-E_{\xi}/T}$. One can then see that Eq.~(\ref{Bol2to1-2}) has a form similar to Eq.~(\ref{Bol1to2-1}),
and  the computation of relic density in this case is also similar to the previous case.
Thus we write the comoving number density for the $A+B \to \xi$ process as
\begin{align}
Y_{\xi} \approx \frac{g_{\xi}}{2\pi^2}\Gamma_{\xi}m^2_{\xi} \int_{T_{\rm min}}^{T_{\rm max}}  \frac{{\rm d}T}{s(T)H(T)} K_1(x_{\xi})\,.
\label{Yx21}
\end{align}
Using Eqs.~(\ref{ST}) and (\ref{HT}), setting $x_{\xi}=m_{\xi}/T$  and carrying out the integration in  Eq.~(\ref{Yx21}) gives
\begin{equation}
Y_{\xi} \approx\frac{135 M_{\rm Pl}}{8\pi^3(1.66)g_{*S}\sqrt{g_*}}\frac{g_{\xi}\Gamma_{\xi}}{m^2_{\xi}}\,.
\end{equation}
Once $\xi$ is produced via $A+B\to\xi$ process, it would subsequently decay to the LSP.
If $A$ is the LSP dark matter particle such as in the process $\tilde\chi^0_1 + h/Z \to \tilde\xi^0_1$,
then using Eq.~(\ref{reliceq})
the contribution to the dark matter relic density from $A+B\to\xi$ process is given by
\begin{equation}
\Omega h^2 \approx\frac{1.1\times 10^{27}}{g_{*S}\sqrt{g_*}}g_{\xi}\frac{m_A\Gamma_{\xi}}{m^2_{\xi}}\,,
\end{equation}
where in our case $g_{\xi}=2$ for the hidden sector neutralinos.
If $A$ is some other bath particle heavier than the dark matter particle with mass $m_{\rm DM}$
such as in the process $\tilde\chi^{\pm}_1 + W^{\mp} \to \tilde\xi^0_1$,
the contribution to the dark matter relic density from $A+B\to\xi$ process is then given by
\begin{equation}
\Omega h^2 \approx \frac{1.1\times 10^{27}}{g_{*S}\sqrt{g_*}}g_{\xi}\frac{m_{\rm DM}\Gamma_{\xi}}{m^2_{\xi}}\,.
\end{equation}

\noindent {\bf{$A+B\to C+\xi$  process:}}

Here two bath particles $A$ and $B$ in thermal equilibrium scatter into $C+\xi$
where $C$ is another bath particle in thermal equilibrium
and $\xi$ as above is the hidden sector particle which has ultraweak interactions with the MSSM sector
and with negligible initial abundance.
 Possible processes of this type are summarized in the Appendix.
For this process the Boltzmann equation for the number density of $\xi$ is given by
\begin{align}
\dot{n}_{\xi}+3Hn_{\xi} & =\int{\rm d}\Pi_{A}{\rm d}\Pi_{B}{\rm d}\Pi_{C}{\rm d}\Pi_{\xi}(2\pi)^{4}\delta^{4}(p_{\xi}+p_{C}-p_{A}-p_{B})\times\nonumber \\
 & \left[|M|_{AB\to C\xi}^{2}f_{A}f_{B}(1\pm f_{C})(1\pm f_{\xi})-|M|_{C\xi\to AB}^{2}f_{\xi}f_{C}(1\pm f_{A})(1\pm f_{B})\right]\nonumber \\
 & \approx\int{\rm d}\Pi_{A}{\rm d}\Pi_{B}{\rm d}\Pi_{C}{\rm d}\Pi_{\xi}(2\pi)^{4}\delta^{4}(p_{\xi}+p_{C}-p_{A}-p_{B})|M|_{AB\to C\xi}^{2}f_{A}f_{B}\,.\label{Bol2to2}
\end{align}
Again $|M|_{AB\to C\xi}^{2}$ is summed over initial and
final spins. Recall that the differential cross-section is given by
\begin{equation}
\sigma_{AB}=\frac{1}{2E_{A}2E_{B}|v_{AB}|}{\rm d}\Pi_{C}{\rm d}\Pi_{\xi}(2\pi)^{4}\delta^{4}(p_{\xi}+p_{C}-p_{A}-p_{B})|\mathcal{M}|_{AB\to C\xi}^{2}\,,
\end{equation}
where $|\mathcal{M}|^{2}$ is averaged over initial spins and summed
over final spins. Notice that
\begin{equation}
{\rm d}\Pi_{C}{\rm d}\Pi_{\xi}(2\pi)^{4}\delta^{4}(p_{\xi}+p_{C}-p_{A}-p_{B})|M|_{AB\to C\xi}^{2}=g_{A}g_{B}\sigma_{AB}v_{AB}2E_{A}2E_{B}\,,
\end{equation}
where
\begin{equation}
v_{AB}=\frac{\sqrt{(p_{A}\cdot p_{B})^{2}-E_{A}^{2}E_{B}^{2}}}{E_{A}E_{B}}\,.
\end{equation}
Thus now Eq. (\ref{Bol2to2}) reduces to~\cite{Edsjo:1997bg}
\begin{align}
\dot{n}_{\xi}+3Hn_{\xi}=&\int{\rm d}\Pi_{A}{\rm d}\Pi_{B}g_{A}g_{B}\sigma_{AB}v_{AB}2E_{A}2E_{B}f_{A}f_{B} \nonumber \\
=&\frac{Tg_{A}g_{B}}{8\pi^{4}}\int_{(m_{A}+m_{B})^{2}}^{\infty}{\rm d}s\ \sqrt{s}\,p_{AB}^{2}\sigma_{AB}K_{1}\left(\frac{\sqrt{s}}{T}\right)\,,
\end{align}
where
\begin{equation}
p_{AB} =\frac{\sqrt{s-(m_{A}+m_{B})^{2}}\sqrt{s-(m_{A}-m_{B})^{2}}}{2\sqrt{s}}
=\frac{v_{AB}E_{A}E_{B}}{\sqrt{s}}\,.
\end{equation}

\section{Model implementation and parameter scan}\label{sec:implement}

\begin{figure}
 \centering
  \subfloat[]{\includegraphics[width=0.49\textwidth]{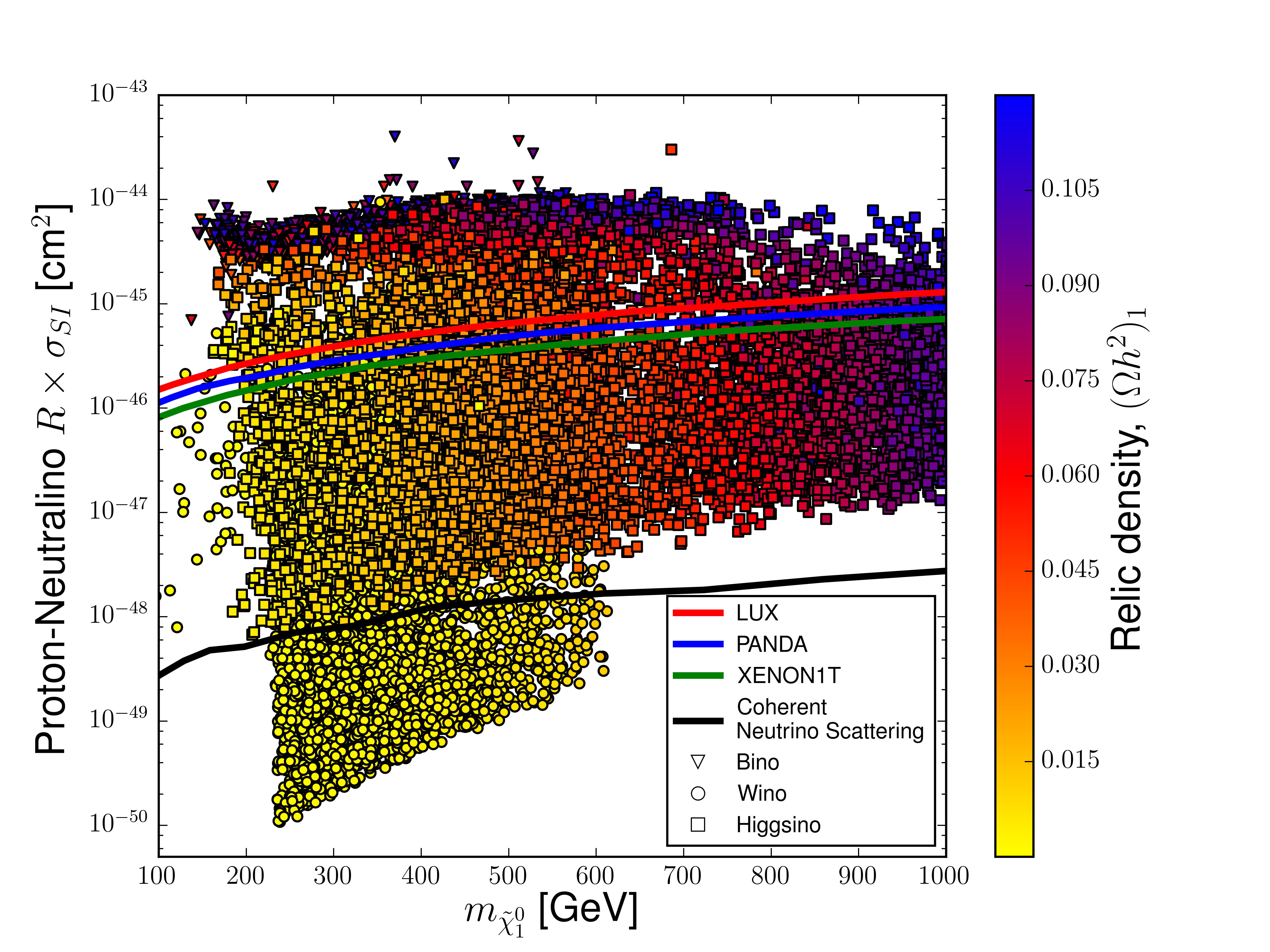}}
  \subfloat[]{\includegraphics[width=0.49\textwidth]{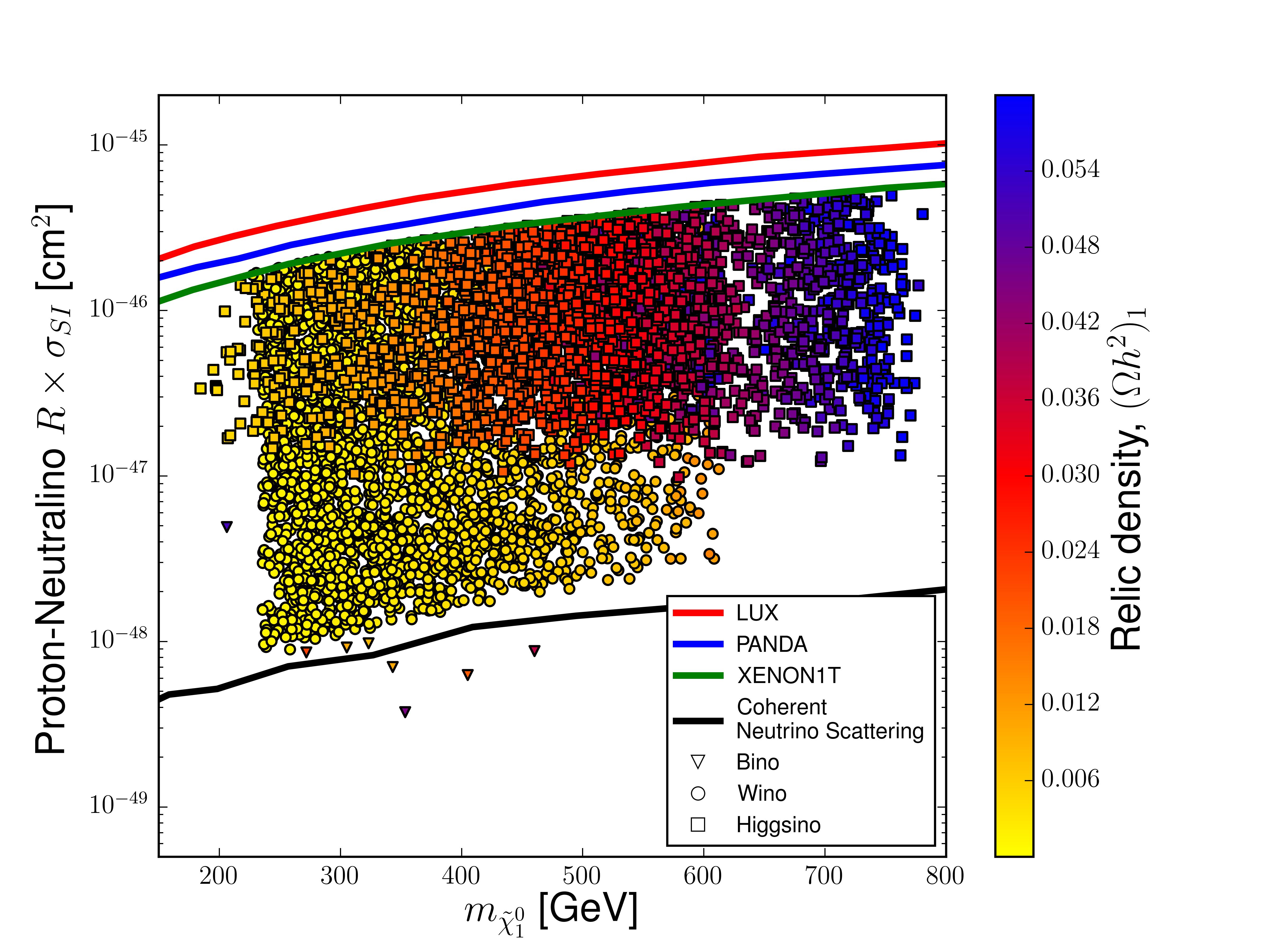}}\\
\subfloat[]{\includegraphics[width=0.49\textwidth]{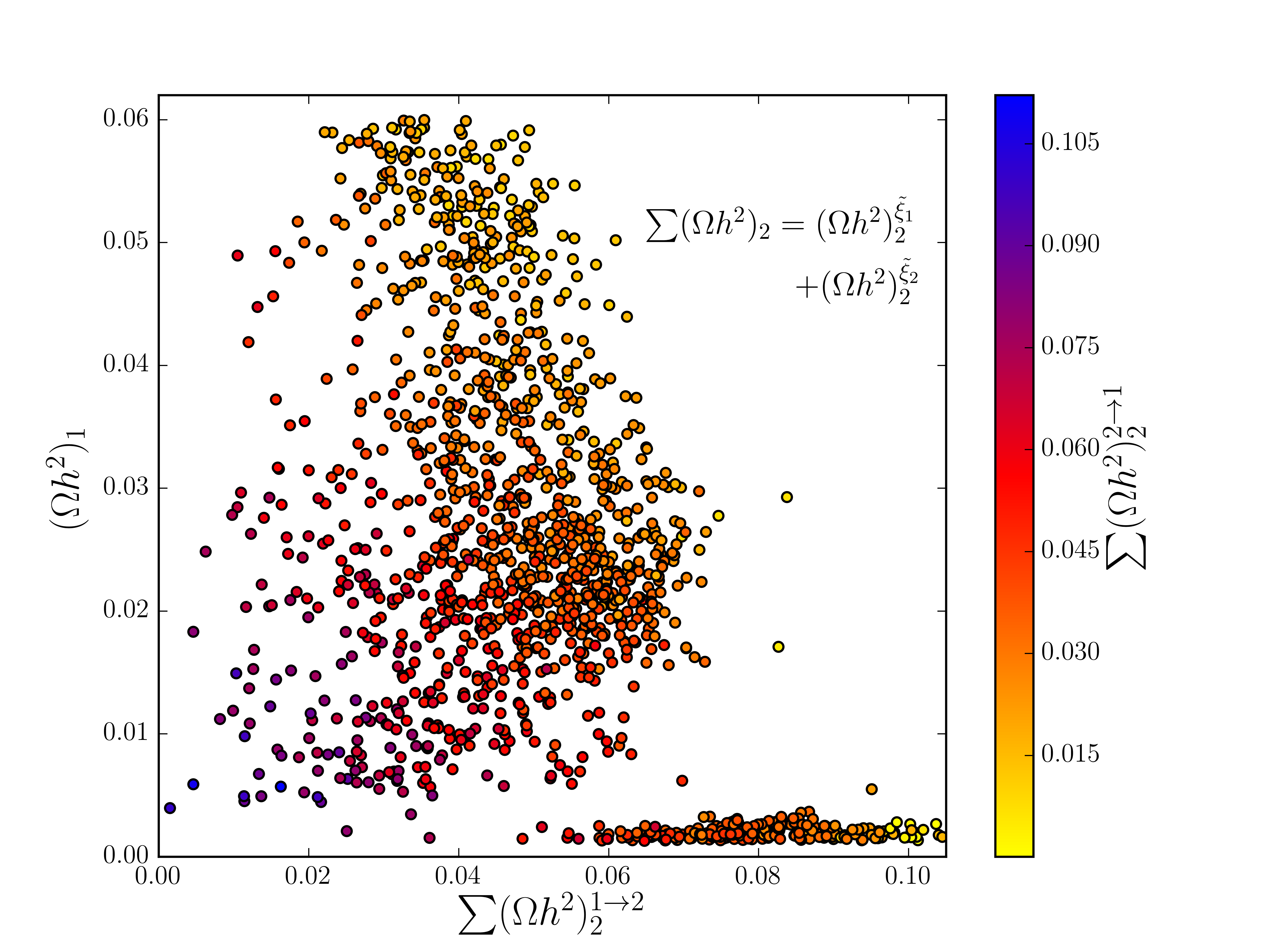}}
\subfloat[]{\includegraphics[width=0.49\textwidth]{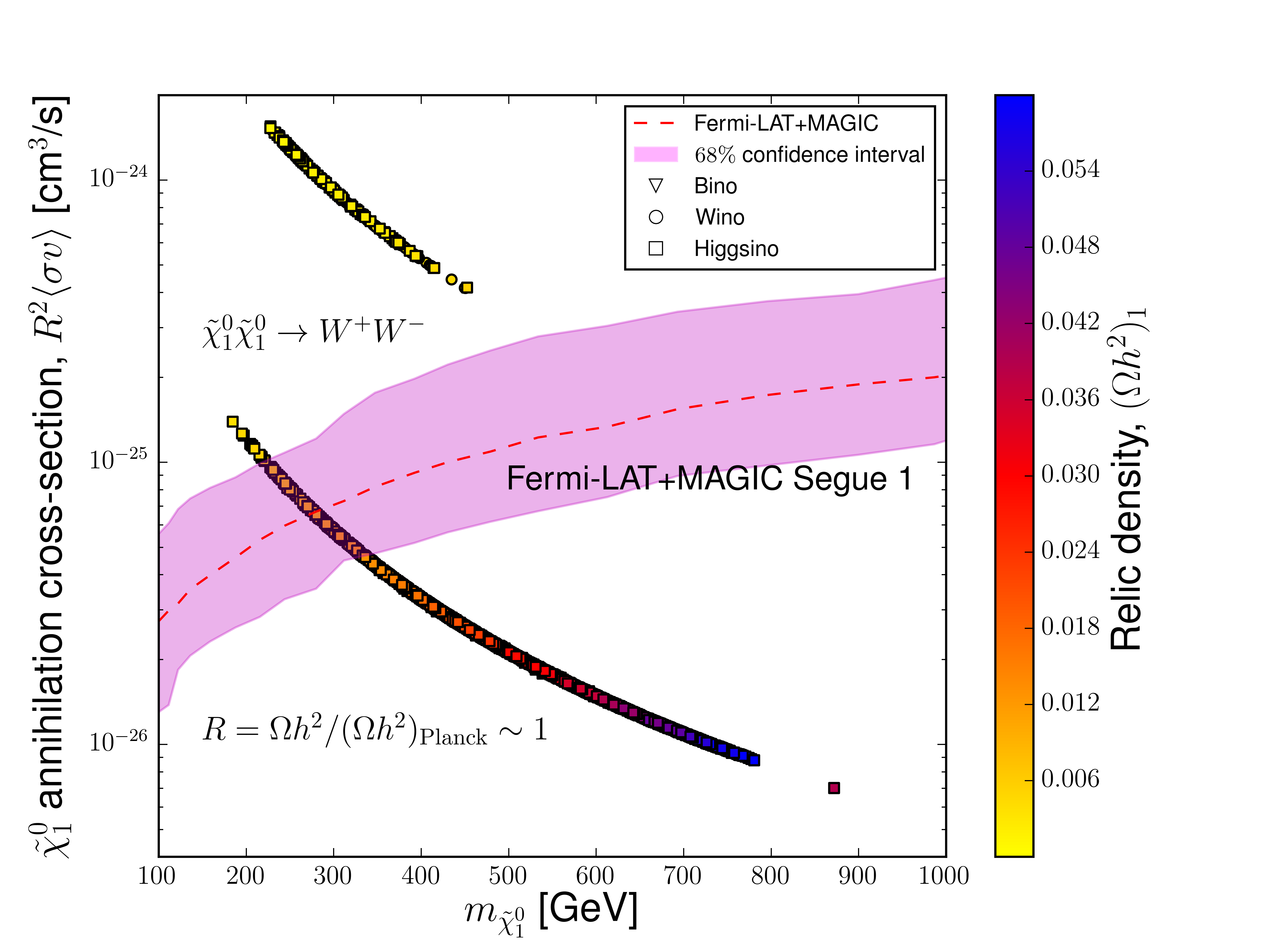}}
   \caption{Panels (i), (ii): Scatter plots of the proton-neutralino spin-independent cross-section, $R\times\sigma_{\rm SI}$, (in cm$^2$) versus the neutralino mass (in GeV), where $R=(\Omega h^2)_{1}/(\Omega h^2)_{\rm PLANCK}$. The bino, wino and higgsino nature of the LSP is exhibited
   by three markers (see legend). Panel (i) shows all points satisfying the Higgs boson mass and $(\Omega h^2)_{\rm 1} < 0.12$ while panel (ii) includes the additional constraint from the XENON1T limit on direct detection  and the contribution to $R$ from the hidden sector ($R=\Omega h^2/(\Omega h^2)_{\rm PLANCK}\sim 1$). Panel (iii): plot shows the different relic density contributions, namely, the freeze-out (along $y$-axis), all $A\to B+\xi$ freeze-in processes ($x$-axis) and the smaller freeze-in contribution from $A+B \to \xi$ processes
  where the relative contributions are indicated by color coding.
   Panel (iv) shows the neutralino annihilation cross-section versus the neutralino mass in the $W^+W^-$ channel with the combined limit from Fermi-LAT and MAGIC experiments
   where the  68\% confidence interval is shown. The colors bar to the right of the panels (i), (ii) and (iv) gives the contribution $(\Omega h^2)_{\rm 1}$ to the freeze-out relic density. Points corresponding to hidden sector neutralinos with lifetimes longer than 10 seconds are removed from panels (iii) and (iv).}
\label{fig1}
\end{figure}

The model described in Sections~\ref{sec:model} and \ref{sec:DM} is implemented with high scale boundary conditions using the mathematica package \code{SARAH-4.14}~\cite{Staub:2013tta,Staub:2015kfa} that generates files for the spectrum generator \code{SPheno-4.0.4}~\cite{Porod:2003um,Porod:2011nf} which runs the two-loop renormalization group equations (RGE) starting from a high scale input taking into account threshold effects to produce the loop-corrected sparticle masses and calculate their decay widths.  \code{SARAH} also generates \code{CalcHep/CompHep}~\cite{Pukhov:2004ca,Boos:1994xb} files used by \code{micrOMEGAs-5.0.9}~\cite{Belanger:2014vza} to determine the dark matter relic density via the freeze-out and freeze-in routines and \code{UFO} files~\cite{Degrande:2011ua} which are input to \code{MadGraph5}~\cite{Alwall:2014hca}.
Our analysis is based on the supergravity grand unified model~\cite{msugra} (for a review see~\cite{Nath:2016qzm}).
The Tadpole equations are solved in terms of $m^2_{H_u}$ and $m^2_{H_d}$, the Higgs soft supersymmetry breaking parameters, and $v_{\rho}$, the VEV developed by the real scalar component of the additional chiral scalar superfield $S$. This method allows us to have $\mu$, the Higgs mixing parameter, as a high scale input of the model. Hence,  the input parameters of the $U(1)_X$-extended MSSM/SUGRA~\cite{msugra,Nath:2016qzm} are of the usual non-universal SUGRA model with additional parameters (all at the GUT scale):
$m_0, ~A_0, ~m_1, ~m_2,  ~m_3, ~\tan\beta, ~\mu, ~\text{sgn}(\mu), ~M_1, ~m_X, ~\delta$,
where $m_0, A_0, m_1, m_2, m_3, \tan\beta$ and $\text{sgn}(\mu)$ are the soft parameters in the MSSM sector as defined earlier.
The parameters $M_2$ and $M_{XY}$ are set to zero at the GUT scale. However, those parameters acquire a tiny value at the electroweak scale due to RGE running. In scanning the parameter space of the model we accept points satisfying the Higgs boson mass and dark matter relic density constraints. Taking theoretical uncertainties into consideration, the constraint of the Higgs mass is taken to be $125\pm 2$ GeV while the relic density is taken in the range 0.110$-$0.128. In generating acceptable parameter points constraints on the sparticle
spectrum implied by the LHC data are also taken into account.
The scan is carried out with \code{xBIT}~\cite{Staub:2019xhl} which uses \code{pytorch} for artificial neural networks (ANN) and \code{xSLHA}~\cite{Staub:2018rih} for reading SLHA files. In the scan, we employ an ANN with three hidden layers and 25 neurons per layer. The result of the scan is shown in Fig.~\ref{fig1}.

In panel (i) of Fig.~\ref{fig1} we show a scatter plot for the proton-neutralino spin-independent cross-section, $R\times\sigma_{\rm SI}$, versus the dark matter mass, with $R=(\Omega h^2)_{\rm 1}/(\Omega h^2)_{\rm PLANCK}$ and $(\Omega h^2)_{\rm PLANCK}$ the measured dark matter relic density by the Planck Collaboration~\cite{Aghanim:2018eyx}
\begin{equation}
(\Omega h^2)_{\rm PLANCK}=0.1198\pm 0.0012\,.
\label{planck}
\end{equation}
The color coding exhibits the dark matter relic density from freeze-out processes (including coannihilation). Many points are already above the current limits from LUX~\cite{Akerib:2016vxi}, PANDA~\cite{Cui:2017nnn} and XENON1T~\cite{Aprile:2018dbl} while others (mostly wino-like neutralinos) are not within reach as they lie below the coherent neutrino scattering floor. Here we do not consider yet contributions to the relic density due to the decay from hidden sector neutralinos. The nomenclature `bino' ($\tilde B$), `wino' ($\tilde W$) and `higgsino' ($\tilde H_u$ and $\tilde H_d$) correspond to the content of the neutralino LSP which can be written as
$\tilde\chi^0_1=\alpha\tilde B+\beta\tilde W+\gamma \tilde H_u+\delta\tilde H_d\,.$
We consider the LSP to be mainly b(w)(higgs)ino if max$(\alpha,\beta,\sqrt{\gamma^2+\delta^2})=\alpha(\beta)(\sqrt{\gamma^2+\delta^2})$. Next we switch on the hidden sector contributions to the relic density via the freeze-in mechanism. Heavy sparticles will decay to the hidden sector neutralino $\tilde\xi^0_1$ and $\tilde\xi^0_2$ which in turn decay to the visible LSP through the ultraweak couplings. Where it exists, the deficit in the neutralino number density (mainly for the wino-like) is made up by the decay of $\tilde\xi^0_1$ and $\tilde\xi^0_2$ raising this number above the neutrino floor. This is exhibited in panel (ii) of Fig.~\ref{fig1} where we see that the  models which in the absence of the hidden sector contribution
were undetectable are now lifted above the neutrino floor and should be within reach of future direct detection experiments.

Now the total relic density is no longer only due to freeze-out  but is given by Eq.~(\ref{totrelic}) which includes the hidden sector contribution and so $R=\Omega h^2/(\Omega h^2)_{\rm PLANCK}\sim 1$. Model points
lying above the XENON1T limit have been eliminated.
A further elimination of model points is applied when $\tilde\xi^0_1$ and $\tilde\xi^0_2$ lifetimes exceed 10 seconds which results in panels (iii) and (iv) of Fig.~\ref{fig1}.
In panel (iii) we exhibit the freeze-out as well as the freeze-in contributions from $A\to B+\xi$  and $A + B\to \xi$  processes
while the total relic density is consistent with Eq.~(\ref{planck}).
In panel (iv) we show the neutralino thermally averaged annihilation cross-section versus the neutralino mass in the dominant $W^+W^-$ channel.
The combined experimental limit from indirect detection experiments, Fermi-LAT and MAGIC Collaborations~\cite{Ahnen:2016qkx} is shown with the 68\% confidence interval.  Model points with a freeze-out relic density less than 0.006 are above the experimental limit (upper branch) while model points with a larger freeze-out relic density and mass greater than $\sim 230$ GeV lie below (or within) the current bound (lower branch).
The points on the upper branch have mostly wino-like LSP and have a  very compressed spectrum with a chargino-LSP mass difference less than 0.3 GeV while the  points on the lower branch have mostly higgsino-like LSP with a less compressed spectrum (mass gap greater than 1 GeV). The proximity of the chargino mass to the LSP mass gives a larger $t$-channel contribution to $\tilde\chi^0_1\tilde\chi^0_1\rightarrow W^+W^-$ for the upper branch due to chargino exchange which also explains the smaller relic density. We note here that all the benchmarks in Table~\ref{tab1}
lie on the lower branch and are consistent with the experimental limits from Fermi-LAT
and MAGIC Collaborations~\cite{Ahnen:2016qkx}.
A further discussion on the compressed spectrum is given in Section~\ref{sec:collider}.

The presence of the ultraweakly coupled hidden sector has expanded the allowed MSSM parameter space with parameter points
which could be detected by future experiments such as XENONnT and LUX-ZEPLIN~\cite{Akerib:2018lyp}.
In the analysis here we have not taken into account the effect of phases to which the neutralino-proton cross sections
are sensitive~\cite{Chattopadhyay:1998wb}.  However, their inclusion would  not significantly affect the conclusions of our analysis.
For a thorough collider study of possible detection of electroweakinos, we proceed by selecting ten benchmarks from panel (iv) after removing the model points lying above the indirect detection limit (the upper branch). In Table~\ref{tab1} we exhibit those points which as we said satisfy all the constraints discussed above. The $\mu$ parameter ranges from $\sim 200$ GeV to $\sim 800$ GeV which is much less than $m_1$ and $m_2$. As a result, all the LSPs in our ten benchmarks are mostly higgsino-like and relatively light.

The sparticle spectrum is displayed in Table~\ref{tab2}. The large $m_0$ and $m_3$ values render the stop and gluino masses heavy while satisfying the Higgs boson mass constraint. The electroweakinos, $\tilde\chi_1^0$, $\tilde\chi_2^0$ and $\tilde\chi_1^\pm$ are relatively light (less than one TeV) with small mass splittings (less than 6 GeV). The reason for this small mass splitting and LHC constraints on the gaugino masses will be discussed in Section~\ref{sec:collider}. We also show in Table~\ref{tab2} the contributions to the relic density from freeze-out and  freeze-in where we see that the dominant contribution to $\Omega h^2$ comes from freeze-in. The masses of the hidden sector neutralinos and the lifetime of $\tilde\xi^0_1$ are also shown.

\begin{table}[H]
\begin{center}
\begin{tabulary}{1.00\textwidth}{l|CCCCCCCCCC}
\hline\hline\rule{0pt}{3ex}
Model & $m_0$ & $A_0$ & $m_1$ & $m_2$ & $m_3$ & $M_1$ & $m_X$ & $\mu$ & $\tan\beta$ & $\delta$ \\
\hline\rule{0pt}{3ex}
\!\!(a) & 4457 & -8315 & 7817 & 5583 & 3595 & 1166 & 2673 & 234 & 6 & 1.6$\times 10^{-12}$ \\
(b) & 7276 & -16268 & 3844 & 3844 & 3844 & 1254 & 1105 & 305 & 37 & 1.2$\times 10^{-12}$ \\
(c) & 439 & 818 & 9725 & 2988 & 5086 & 1667 & 2777 & 354 & 30 & 1.8$\times 10^{-12}$ \\
(d) & 5202 & -5343 & 7332 & 7332 & 5735 & 1778 & 1087 & 416 & 9 & 2.1$\times 10^{-12}$ \\
(e) & 943 & -1331 & 11670 & 2523 & 3531 & 1387 & 406 & 515 & 10 & 2.3$\times 10^{-12}$ \\
(f) & 4677 & -2052 & 11837 & 5940 & 5822 & 2278 & 1087 & 597 & 35 & 2.8$\times 10^{-12}$ \\
(g) & 1164 & -1061 & 4667 & 4667 & 5038 & 1558 & 653 & 640 & 32 & 5.0$\times 10^{-13}$ \\
(h) & 2382 & -2881 & 2939 & 2939 & 5110 & 1735 & 425 & 671 & 12 & 3.4$\times 10^{-12}$ \\
(i) & 5796 & -13224 & 7363 & 7363 & 6849 & 1296 & 1074 & 682 & 5 & 1.3$\times 10^{-12}$ \\
(j) & 2030 & -759 & 2971 & 2971 & 2360 & 1699 & 366 & 865 & 29 & 2.7$\times 10^{-12}$ \\
\hline
\end{tabulary}\end{center}
\caption{Input parameters for the benchmarks used in this analysis. Here $M_2=M_{XY}=0$ at the GUT scale. All masses are in GeV.}
\label{tab1}
\end{table}

\begin{table}[H]
\begin{center}
\resizebox{1\linewidth}{!}{\begin{tabulary}{\textwidth}{l|cccccccccccc}
\hline\hline\rule{0pt}{3ex}
Model  & $h^0$ & $\tilde\chi_1^0$& $\tilde\chi_2^0$ & $\tilde\chi_1^\pm$ &  $\tilde{\xi}^0_1$ & $\tilde{\xi}^0_2$ & $\tilde t$ & $\tilde g$ & $(\Omega h^2)_{\rm 1}$ & $ (\Omega h^2)_{2}$ & $\Omega h^2$ & $\tau_0$ \\
\hline\rule{0pt}{3ex}
\!\!(a) & 124.6 & 251.3 & 253.3 & 252.5 & 437 & 3110 & 4016 & 7423 & 0.007 & 0.100 & 0.107 & 1.95 \\
(b) & 124.0 & 301.1 & 304.4 & 303.1 & 818 & 1923 & 2486 & 8016 & 0.010 & 0.095 & 0.105 & 0.51 \\
(c) & 125.2 & 364.2 & 367.2 & 365.9 & 781 & 3558 & 7289 & 10036 & 0.014 & 0.100 & 0.115 & 0.35 \\
(d) & 125.7 & 450.6 & 452.3 & 451.8 & 1316 & 2403 & 7713 & 11389 & 0.021 & 0.092 & 0.113 & 0.09 \\
(e) & 123.8 & 551.5 & 555.3 & 553.1 & 1199 & 1605 & 5324 & 7153 & 0.031 & 0.079 & 0.110 & 0.07\\
(f) & 126.2 & 601.7 & 603.4 & 602.9 & 1798 & 2885 & 8497 & 11515 & 0.037 & 0.086 & 0.123 & 0.03 \\
(g) & 125.4 & 649.6 & 652.4 & 651.4 & 1265 & 1918 & 6734 & 9912 & 0.044 & 0.078 & 0.122 & 1.64\\
(h) & 125.4 & 717.1 & 722.4 & 720.3 & 1535 & 1960 & 6851 & 10098 & 0.053 & 0.071 & 0.124 & 0.04 \\
(i) & 124.9 & 730.0 & 731.8 & 731.2 & 866 & 1940 & 7253 & 13410 & 0.054 & 0.054 & 0.108 & 2.13 \\
(j) & 123.0 & 872.8 & 878.6 & 876.5 & 1526 & 1892 & 3502 & 4940 & 0.039 & 0.084 & 0.123 & 0.06 \\
\hline
\end{tabulary}}\end{center}
\caption{Display of the Higgs boson ($h^0$) mass, the stop ($\tilde t$) mass, the
  gluino ($\tilde g$)  mass,  the relevant electroweakino ($\tilde \chi_1^0, \tilde \chi_2^0,\tilde \chi^{\pm}_1$) masses,
  the hidden sector neutralino $\xi_1^0,\xi_2^0$ masses,
  and the relic density for the benchmarks  of Table~\ref{tab1} computed at the electroweak scale.
  $\tau_0$ (in s) is the lifetime of the
  hidden sector neutralino $\tilde\xi^0_1$ which has a decay consistent with the BBN constraint. All masses are in GeV. }
\label{tab2}
\end{table}

For the benchmarks of Table~\ref{tab1}, we display in Table~\ref{tabxs} the spin-independent proton-neutralino scattering cross-section and the thermally averaged neutralino annihilation cross-section satisfying the bounds from XENON1T for direct detection and Fermi-LAT and MAGIC for indirect detection in the $W^+W^-$ channel.

\begin{table}[H]
\begin{center}
\begin{tabulary}{1.05\textwidth}{l|CC}
\hline\hline\rule{0pt}{3ex}
Model & proton-$\tilde\chi^0_1$ cross-section, & $\tilde\chi^0_1$ annihilation cross-section, \\
  & $R\times\sigma_{\rm SI}$ [cm$^2$]  &  $R^2\langle\sigma v\rangle_{W^+W^-}$ [cm$^3$/s] \\
\hline\rule{0pt}{3ex}
\!\!(a) & $5.33\times 10^{-47}$ & $8.04\times 10^{-26}$  \\
(b) & $9.18\times 10^{-47}$ & $5.69\times 10^{-26}$    \\
(c) & $1.13\times 10^{-46}$ & $3.96\times 10^{-26}$   \\
(d) & $3.16\times 10^{-47}$ & $2.61\times 10^{-26}$  \\
(e) & $2.35\times 10^{-46}$ & $1.76\times 10^{-26}$    \\
(f) & $2.48\times 10^{-47}$ & $1.49\times 10^{-26}$    \\
(g) & $8.12\times 10^{-47}$ & $1.26\times 10^{-26}$    \\
(h) & $4.24\times 10^{-46}$ & $1.04\times 10^{-26}$    \\
(i) & $4.27\times 10^{-47}$ & $1.01\times 10^{-26}$    \\
(j) & $6.04\times 10^{-46}$ & $7.01\times 10^{-27}$   \\
\hline
\end{tabulary}\end{center}
\caption{The spin-independent proton-neutralino scattering cross-section, $R\times\sigma_{\rm SI}$ (second column) and the thermally averaged neutralino annihilation cross-section, $R^2\langle\sigma v\rangle_{W^+W^-}$ in the $W^+W^-$ channel (third column) for the ten benchmarks of Table~\ref{tab1}. Here $R\sim 1$ due to the hidden sector contribution. }
\label{tabxs}
\end{table}

\section{Collider study of a compressed electroweakino spectrum}\label{sec:collider}
Models of natural supersymmetry requiring small $\mu$ are highly constrained by the LEP and LHC data.  However, regions of parameter space exist
consistent with the current experimental limits where models with relatively small $\mu$ lead to electroweakino masses which would be accessible for
discovery at HL-LHC and HE-LHC. We discuss here a class of models with these characteristics as given in Table 1 and Table 2. One
characteristic of these models is that the electroweakino mass spectrum is compressed with
the chargino-lightest neutralino mass gap ranging  from $\sim 1$ to $\sim 4$ GeV. The hierarchy between $m_1$, $m_2$ and $\mu$ determines how much compressed the spectrum is. We distinguish here between two cases: $m_1\gg m_2>\mu$ and $m_2\gg m_1>\mu$ where we  have taken the sign of $\mu$ to be positive.

\paragraph{Case 1: $m_1\gg m_2>\mu$}
Here we consider the $4\times 4$ MSSM neutralino mass matrix with small $\mu$ for the case $m_Z^2\ll|m_{1,2}\pm \mu|^2$, where $m_Z$ is the $Z$ boson
mass where the lightest neutralinos are higgsino-like and their masses are given by~\cite{Choi:2001ww}
\begin{align}
m_{\tilde\chi^0_1}=\mu-\left(\frac{1+\sin 2\beta}{2}\right)\left[(K_1+K_2)\mu+K_1m_1+K_2m_2\right], \nonumber \\
m_{\tilde\chi^0_2}=\mu-\left(\frac{1-\sin 2\beta}{2}\right)\left[(K_1+K_2)\mu-K_1m_1-K_2m_2\right],
\end{align}
where
\begin{equation}
K_1=\frac{m_Z^2\sin^2\theta_W}{m_1^2-\mu^2}, ~~~\text{and}~~~ K_2=\frac{m_Z^2\cos^2\theta_W}{m_2^2-\mu^2},
\end{equation}
where $\theta_W$ the weak mixing angle and $\mu$, $\beta$, $m_1$ and $m_2$  assume their values at the electroweak scale. The lightest chargino which is mostly Higgsino has a mass
\begin{equation}
m_{\chi^{\pm}_1}=\mu-K_2(\mu+m_2\sin 2\beta).
\end{equation}
In this case, the chargino-LSP and the second neutralino-LSP mass differences are given by
\begin{align}
\Delta m_1=m_{\chi^{\pm}_1}-m_{\tilde\chi^0_1}\approx \frac{m^2_W(1-\sin 2\beta)}{2(m_2+\mu)}, \nonumber \\
\Delta m_2=m_{\chi^{0}_2}-m_{\tilde\chi^0_1}\approx \frac{m^2_W(\mu\sin 2\beta+m_2)}{m_2^2-\mu^2},
\label{deltam1}
\end{align}
where $m_W$ is the $W$ boson mass. We note that the chargino-neutralino  masses
 become more degenerate the larger $m_2$ is. In particular this is true for benchmarks (c) and (e) of Table~\ref{tab1} where the mass gap is $\sim 1.6$ GeV (see the spectrum in Table~\ref{tab2}).

\paragraph{Case 2: $m_2\gg m_1>\mu$}
Here the heavy wino component can be integrated out and the mass difference between the light chargino and lightest neutralino and the second neutralino and the lightest one are given by
\begin{align}
\Delta m_1=m_{\chi^{\pm}_1}-m_{\tilde\chi^0_1}\approx \frac{m^2_W\tan^2\theta_W(1+\sin 2\beta)}{2(m_1-\mu)}, \nonumber \\
\Delta m_2=m_{\chi^{0}_2}-m_{\tilde\chi^0_1}\approx \frac{m^2_W\tan^2\theta_W(\mu\sin 2\beta+m_1)}{m_1^2-\mu^2}.
\label{deltam2}
\end{align}
This case applies in particular to the benchmarks (b), (d), (i) and (j). Note that even if $m_1=m_2$ at the GUT scale (as given in Table~\ref{tab1}), RGE running of the gaugino parameters produces very different values of $m_1$ and $m_2$ at the low scale.
We have seen that by requiring a small $\mu$ and satisfying the LHC constraints on electroweakino masses, we are lead to a compressed spectrum due to
the large $m_1$ and $m_2$ as evident from Eqs.~(\ref{deltam1}) and~(\ref{deltam2}).

Experiments at ATLAS and CMS have set stringent limits on chargino and neutralino masses corresponding to large mass splittings. Chargino mass up to 1.1 TeV and a neutralino of mass $\sim 600$ GeV have been ruled out~\cite{Aad:2015eda}. As for tiny mass splittings, charginos and neutralinos of masses less than 200 GeV have been excluded~\cite{Aaboud:2018jiw,Sirunyan:2018ubx}. Searches targetting mass splittings near the electroweak scale~\cite{Aad:2019vvi,Aaboud:2018sua,Sirunyan:2017lae}, where $W$ and $Z$ bosons are on their mass shell, have led ATLAS to exclude neutralinos and charginos up to 345 GeV while CMS pushed the limit to 475 GeV. Most recently and using 139$\ifb$ of data, ATLAS performed a search for electroweakino pair production for mass splittings of 1.5 GeV to 2.4 GeV~\cite{Aad:2019qnd} where limits on chargino mass have been set at 92 GeV to $\sim 190$ GeV and at 240 GeV for $\Delta m_1=7$ GeV. In this section, we perform a collider analysis study for our benchmarks at HL-LHC and HE-LHC which are characterized by a very small mass splittings ranging from 1.2 GeV to 3.7 GeV for $\Delta m_1$ and up to $\sim 6$ GeV for $\Delta m_2$. This is a very challenging search due to the softness of the final states but it is expected that HL-LHC will have a better electron and muon track reconstruction efficiency even for large pseudorapidity ranges and small lepton transverse momenta (down to 2 or 3 GeV). The replacement of the inner detector in both ATLAS and CMS will extend the coverage to $|\eta|<4.0$ and for the range $2.5<|\eta|<4.0$ the electron efficiency can be $\sim 15\%$ for tight identification requirement and up to $\sim 40\%$ for loose identification requirement at $p_T=5$ GeV~\cite{HL-performance}.

\subsection{Signal and background simulation and LHC production of electroweakino pairs}\label{sec:simulation}

The signal under study consists of a second neutralino production in association with a chargino ($\tilde\chi^0_2\tilde\chi^{\pm}_1$) and a chargino pair production ($\tilde\chi^+_1\tilde\chi^-_1$) with dileptonic final states as shown in Fig.~\ref{fig2}. The leptons may come from the decay of a second neutralino via $Z^*$ (left Feynman diagram) or from two charginos via $W^*$ decay (right Feynman diagram).
Thus the final states we are looking for in this study are at least two soft leptons, jets and a large missing transverse energy due to the neutralino (and neutrinos).
Because of the soft final states, an initial state radiation (ISR)-assisted topology is employed which can boost the sparticle system giving the final states an additional transverse momenta essential for their detection. We present in Table~\ref{tab3} the decay branching ratios (BR) of the second neutralino and the chargino into the final states considered in Fig.~\ref{fig2}. The leptonic channel BR of the second neutralino is $\sim 10\%$ and that of the chargino is close to 40\%. For larger mass gaps, this BR decreases due to the opening of the tau decay channel
[for example for benchmark (j)]. The hadronic decay channel of the chargino is dominant across all benchmarks.

\begin{figure}[H]
 \centering
   \includegraphics[width=0.49\textwidth]{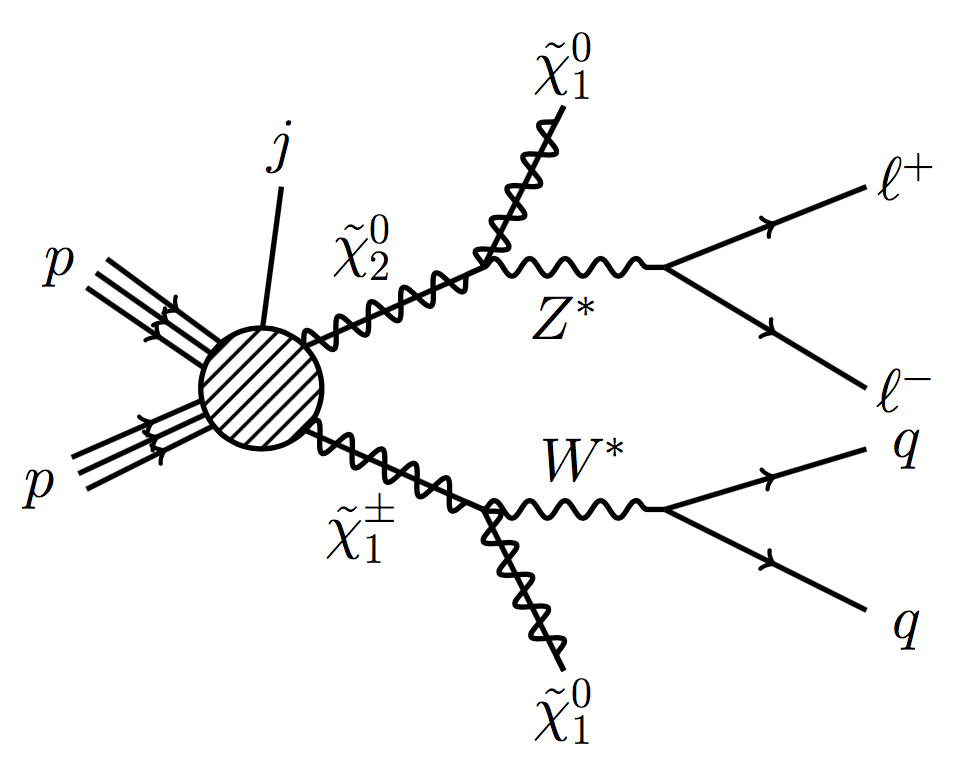}
   \includegraphics[width=0.49\textwidth]{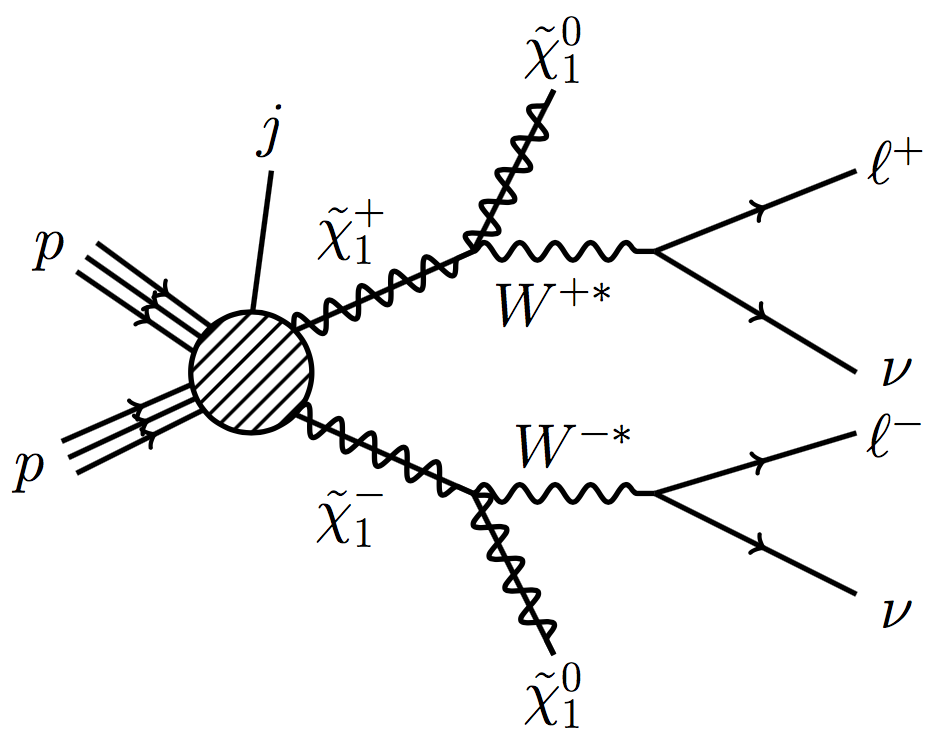}
   \caption{Direct electroweakino pair production in $pp$ collision with leptonic final states due to off-shell $W^*$ and $Z^*$ decays. The line labeled $j$ denotes an ISR jet. }
	\label{fig2}
\end{figure}

\begin{table}[H]
\begin{center}
\begin{tabulary}{1.05\textwidth}{l|CCC}
\hline\hline\rule{0pt}{3ex}
Model & BR$(\tilde\chi^0_2\rightarrow \tilde\chi^0_1\ell^+\ell^-)$ & BR$(\tilde\chi^{\pm}_1\rightarrow \tilde\chi^0_1q_i \bar{q}_j)$ & BR$(\tilde\chi^{\pm}_1\rightarrow \tilde\chi^0_1\ell^{\pm}\nu)$ \\
\hline\rule{0pt}{3ex}
\!\!(a) & 0.098 & 0.604 & 0.396 \\
(b) & 0.095 & 0.624 & 0.374  \\
(c) & 0.095 & 0.608 & 0.395  \\
(d) & 0.099 & 0.604 & 0.396 \\
(e) & 0.084 & 0.607 & 0.392  \\
(f) & 0.099 & 0.604 & 0.396  \\
(g) & 0.099 & 0.613 & 0.385  \\
(h) & 0.087 & 0.666 & 0.303  \\
(i) & 0.099 & 0.604 & 0.396  \\
(j) & 0.086 & 0.669 & 0.284  \\
\hline
\end{tabulary}\end{center}
\caption{The electroweakino branching ratios into the final states shown in Fig.~\ref{fig2}. In the table header, $q_i \bar{q}_j\in\{u\bar{d},u\bar{s},c\bar{d},c\bar{s}\}$ for $\tilde\chi^+_1$ decay and the conjugate of that set for $\tilde\chi^-_1$ decay and $\ell$ denotes electrons and muons.}
\label{tab3}
\end{table}

The production cross-sections of $\tilde\chi^0_2\tilde\chi^{\pm}_1$ and $\tilde\chi^+_1\tilde\chi^-_1$ at next-to-leading order (NLO) in QCD with next-to-next-to-leading logarithm resummation (NNLL) at 14 TeV and 27 TeV are calculated with \code{Resummino-2.0.1}~\cite{Debove:2011xj,Fuks:2013vua} using the five-flavour NNPDF23NLO PDF set. The NLO+NNLL cross-sections for the ten benchmarks of Table~\ref{tab1} are shown below in Table~\ref{tab4}.

\begin{table}[H]
\begin{center}
\begin{tabulary}{1.2\textwidth}{l|cc|cc}
\hline\hline\rule{0pt}{3ex}
Model  & \multicolumn{2}{c}{$\sigma_{\rm NLO+NNLL}(pp\rightarrow \tilde\chi^0_2\,\tilde\chi^{\pm}_1)$} & \multicolumn{2}{c}{$\sigma_{\rm NLO+NNLL}(pp\rightarrow \tilde\chi^+_1\,\tilde\chi^-_1)$} \\
&  &  &  &  \\
  & 14 TeV & 27 TeV & 14 TeV & 27 TeV  \\
\hline\rule{0pt}{3ex}
\!\!(a) & 68.36 & 195.43 & 117.64 & 319.06 \\
(b) & 33.32 & 102.04 & 58.85 & 169.61 \\
(c) & 15.38 & 51.21 & 28.02 & 87.13 \\
(d) & 6.23 & 23.25 & 11.68 & 40.27 \\
(e) & 2.46 & 10.57 & 4.83 & 18.84 \\
(f) & 1.63 & 7.48 & 3.23 & 13.42 \\
(g) & 1.11 & 5.46 & 2.24 & 9.90 \\
(h) & 0.66 & 3.58 & 1.37 & 6.61 \\
(i) & 0.61 & 3.36 & 1.27 & 6.19 \\
(j) & 0.23 & 1.54 & 0.49 & 2.91 \\
\hline
\end{tabulary}
\end{center}
\caption{The NLO+NNLL production cross-sections, in fb, of electroweakinos: the second neutralino-chargino pair, $\tilde\chi^0_2\,\tilde\chi^{\pm}_1$ (second and third columns), and opposite sign chargino pair (fourth and fifth columns) at $\sqrt{s}=14$ TeV and at $\sqrt{s}=27$ TeV for benchmarks of Table~\ref{tab1}.}
\label{tab4}
\end{table}

For the final states, the dominant SM backgrounds are $W/Z/\gamma^*+$ jets, diboson production, $t\bar{t}$, $t+W/Z$ and dilepton production from off-shell vector bosons ($V^*\rightarrow \ell\ell$). The signal and background events are simulated at LO with up to two partons at generator level using \code{MadGraph5\_aMC@NLO-2.6.3} interfaced to \code{LHAPDF}~\cite{Buckley:2014ana} using the NNPDF30LO PDF set. The signal and background cross-sections are then scaled to their NLO+NNLL and NLO values, respectively, at 14 TeV and at 27 TeV. The showering and hadronization of parton level events is done with \code{PYTHIA8}~\cite{Sjostrand:2014zea} using a five-flavour MLM matching~\cite{Mangano:2006rw} in order to avoid double counting of jets. For the signal samples, a matching/merging scale is set at one-fourth the mass of the chargino. Jets are clustered with \code{FastJet}~\cite{Cacciari:2011ma} using the anti-$k_t$ algorithm~\cite{Cacciari:2008gp} with jet radius $R=0.4$. Detector simulation and event reconstruction is handled by \code{DELPHES-3.4.2}~\cite{deFavereau:2013fsa} using the beta card for HL-LHC and HE-LHC studies which addresses the improvements in lepton reconstruction efficiencies. We do not modify those settings which seem reasonably close to what the experimental collaborations are suggesting. Accordingly, an electron reconstruction efficiency for $p_T>4$ GeV ranges from $\sim 35\%$ to $\sim 65\%$ depending on the $\eta$ region whereas the muon reconstruction efficiency can have values starting at $16\%$ for $p_T>2$ GeV in $1.0<|\eta|<1.5$ range. The analysis of the resulting event files and cut implementation is carried out with \code{ROOT 6}~\cite{Antcheva:2011zz}.

\subsection{Analysis technique and event preselection}

For such a highly compressed spectrum and very soft final states, a traditional cut-and-count analysis is inefficient and may lead to maximal loss of the signal relative to the overwhelming SM background. In order to efficiently exploit the ISR-boosted signal topology we employ the recursive jigsaw reconstruction (RJR) technique~\cite{Jackson:2016mfb,Jackson:2017gcy}. The idea is to build a decay tree which describes the signal topology of interest. Each element of this tree behaves as a reference frame of its own where reconstructed objects are assigned to. We show in Fig.~\ref{fig3} the generic decay tree used for compressed spectra. CM stands for the center-of-mass frame from which an ISR jet and a sparticle (S) system arise. The (S) system recoiling against ISR then decays to two categories of states: visible and invisible (I). The latter corresponds to massive LSPs and/or neutrinos. Here we distinguish between two visible states: (J) which contains all jets that are not identified as ISR and (L) which contains reconstructed leptons (electrons and muons).

\begin{figure}[H]
 \centering
   \includegraphics[width=0.49\textwidth]{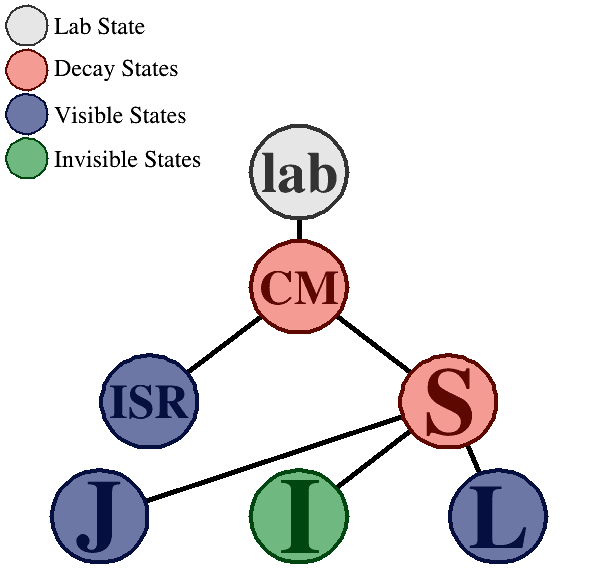}
   \caption{The general compressed decay tree used in the RJR technique. CM denotes the center-of-mass frame giving rise to a back-to-back ISR and (S) systems. The (S) system decays to visible jets (J) and invisible (I) states and to leptons (L).}
	\label{fig3}
\end{figure}

In case where very little transverse momentum is imparted to the LSP due to the very compressed phase space, the missing transverse energy of the system is entirely due to the recoil against ISR and is given by
\begin{equation}
\vec{E}^{\rm miss}_T\sim \vec{p}^{\ \rm ISR}_T\frac{m_{\tilde\chi^0_1}}{M},
\end{equation}
where $M$ is the mass of the parent supersymmetric particle. This relation is only an approximation and assumes that the ISR system is entirely due to a single jet. So it is important to be able to properly calculate $E^{\rm miss}_T/|\vec{p}^{\ \rm ISR}_T|$ for more complicated situations. The RJR technique does exactly that by applying a set of ``jigsaw rules'' which result in a number of observables evaluated in specific reference frames. The rules followed for the reconstruction of events are:
\begin{enumerate}
\item Setting the longitudinal components to zero and considering only the transverse ones.
\item The mass of the invisible system is set to zero.
\item All missing transverse energy is set to the (I) system and reconstructed leptons' four-momenta assigned to the (L) system.
\item The object partitioning between ISR and (J) systems aims at distinguishing ISR jets from jets resulting from the (S) system decay.
\end{enumerate}

The partitioning between ISR and (J) is done by minimizing the reconstructed masses of the (S) system, $m_S$, and the ISR system, $m_{\rm ISR}$. Boosting to the transverse CM frame, we can write the CM mass as
\begin{equation}
m_{\rm CM}=\sqrt{m^2_S+p^{\rm ISR^2}_T}+\sqrt{m^2_{\rm ISR}+p^{\rm S^2}_T},
\end{equation}
where $p^{\rm ISR}$ and $p^{\rm S}_T$ are the transverse momenta of the ISR and (S) systems, respectively, evaluated in the CM frame. With $m_{\rm CM}$ fixed, jets are assigned in such a way to maximize $p^{\rm ISR/S}_T$ with each partitioning of indistinguishable objects into either (J) or ISR systems thereby minimizing the masses of the said systems. The result of applying the above ``jigsaw rules'' is a set of observables that act as a powerful discriminant between the signal and the background. We list the relevant observables hereafter:
\begin{enumerate}
\item The ratio $R_{\rm ISR}$ defined as
\begin{equation}
R_{\rm ISR}=\frac{|\vec{p}^{\ \rm I,CM}_T\cdot\hat{p}^{\,\rm ISR,CM}_T|}{\vec{p}^{\ \rm ISR,CM}_T},
\end{equation}
where $\vec{p}^{\ \rm I,CM}_T$ is the transverse momentum of the invisible system in the CM frame. Here the label `CM' is shown explicitly on all vectors to denote that the variables are determined in the CM frame. In the limit of very small mass splittings, we can approximate this ratio by
\begin{equation}
R_{\rm ISR}\sim\frac{|\vec{E}^{\rm miss}_T\cdot\hat{p}^{\,\rm ISR,CM}_T|}{p^{\rm ISR,CM}_T}\sim\frac{m_{\tilde\chi^0_1}}{M}+\cdots,
\end{equation}
where $\cdots$ correspond to terms which on the average can be taken as zero. One can see that this ratio should peak close to one for the signal. This is exhibited in the left panel of Fig.~\ref{fig4}.

\item $p^{\rm ISR,CM}_T$: the magnitude of the transverse momenta of all ISR jets in an event, evaluated in the CM frame.

\item $N_j^{\rm ISR}$ and $N_j^{\rm V}$: the number of ISR jets and number of `visible' jets from the decay of sparticles, respectively.

\item $\Delta\phi(\rm I, ISR)$: the angle between the ISR system and the invisible system evaluated in the CM frame

\end{enumerate}
We impose some preselection criteria on the signal samples and SM backgrounds before we begin our analysis using the RJR technique and the observables listed above. As mentioned before, the signal region (SR) consists of two leptons, at least one jet and missing transverse energy in the final state. Events are selected with $p_T^{\rm leading~jet}>30$ GeV, lepton tracks with $p_T^{\ell}>4$ GeV and $E^{\rm miss}_T>90$ GeV. A veto is applied on b-tagged and tau-tagged jets which reduces the $t\bar{t}$ background. Another important preselection criteria applies to the dilepton invariant mass, $m_{\ell\ell}$, in case of same flavour and opposite sign (SFOS) leptons. The distribution in $m_{\ell\ell}$ is shown in the right panel of Fig.~\ref{fig4}. The dominant background is from $Z+$ jets with a peak near the $Z$ boson mass. The signal, however, has a much smaller $m_{\ell\ell}$ with most of the events lying in the region $m_{\ell\ell}<5$ GeV which is the characteristic mass splitting in the signal. The larger $m_{\ell\ell}$ values are due to the fact that some SFOS leptons can come from two $W$ bosons on opposite sides of the decay tree, mainly from chargino pair production. Only events in the range $0<m_{\ell\ell}\leq 20$ GeV are accepted which removes a large part of the SM background and retains most of the signal. Note that this is a minimal preselection criteria applied to the HL-LHC analysis. The preselection criteria is slightly modified for the HE-LHC case and is shown in Table~\ref{tab5}.

\begin{figure}[H]
 \centering
   \includegraphics[width=0.49\textwidth]{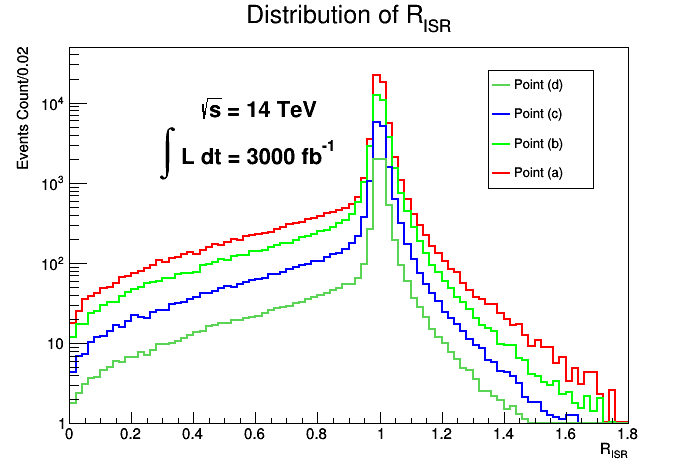}
   \includegraphics[width=0.49\textwidth]{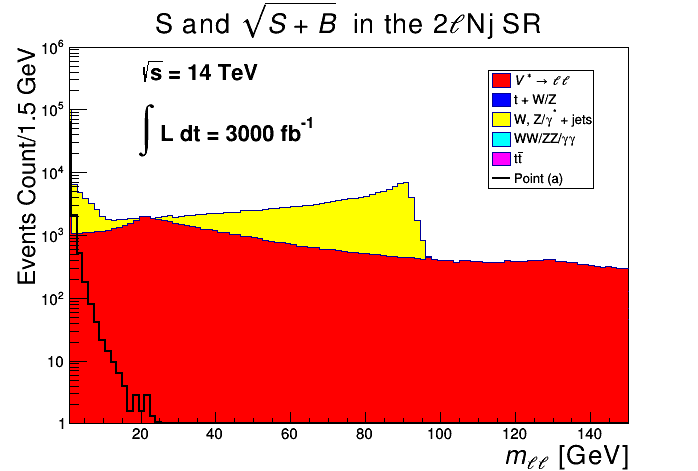}
   \caption{Left panel: distributions in the variable $R_{\rm ISR}$ for four benchmarks (a), (b), (c) and (d) at 14 TeV and 3000$\ifb$ of integrated luminosity. Right panel: distribution in the dilepton invariant mass for the SM background and benchmark (a).}
	\label{fig4}
\end{figure}

\subsection{Selection criteria and results}\label{sec:results}

In addition to the RJR observables derived in the previous section we will use a few more variables which will help us discriminate the signal from the SM background:
\begin{enumerate}
\item $E^{\rm miss}_T/H^{\rm lep}_T$: here $H^{\rm lep}_T$ is the sum of the transverse momenta of the first two leading leptons. This is a very effective variable since the signal is characterized by a large missing tranverse energy and soft leptons.

\item The di-tau invariant mass, $m_{\tau\tau}$, is very effective in rejecting $Z/\gamma^*\rightarrow \tau\tau~+$ jets background~\cite{Han:2014kaa,Baer:2014kya,Barr:2015eva}. In order to calculate this variable, we consider the leptonic decays of the tau and assume that taus are highly relativistic. This means that their leptonic products and neutrinos are almost collinear with each other and moving in the same direction as the parent tau. With this in mind, the total missing transverse momentum due to the neutrinos can be written as
\begin{equation}
\vec{p}_T^{\ \rm miss}=\kappa_1\vec{p}_T^{\ \ell_1}+\kappa_2\vec{p}_T^{\ \ell_2}.
\end{equation}
This is basically a set of two independent equations which can be solved to determine $\kappa_1$ and $\kappa_2$ leading to an estimate of $m_{\tau\tau}^2$ determined by
\begin{equation}
m_{\tau\tau}^2=2(1+\kappa_1)(1+\kappa_2)m^2_{\ell\ell}.
\end{equation}
The quantities $\kappa_1$ and $\kappa_2$ can assume negative values if $p_T^{\ell}$ is smaller than $E^{\rm miss}_T$ and points in a different direction to $\vec{p}_T^{\ \rm miss}$. This can happen if neutrinos coming from the decay of SM particles are paired with leptons of uncorrelated directions. So one can see that $m_{\tau\tau}^2$ can be negative and so the di-tau invariant mass is determined as $m_{\tau\tau}=\text{sign}(m_{\tau\tau}^2)\sqrt{|m_{\tau\tau}^2|}$.
\end{enumerate}

In Fig.~\ref{fig5} we exhibit distributions of four kinematic variables for points (a) [upper panels (i) and (ii)], (b) and (c) [lower panels (iii) and (iv)] at 14 TeV and 3000$\ifb$ of integrated luminosity after applying the preslection criteria. For smaller values of the variable $E^{\rm miss}_T/H^{\rm lep}_T$, the SM background shows a large increase unlike the signal which makes it an effective variable in eliminating a large part of the background. Even though an excess of signal events is clear for $E^{\rm miss}_T/H^{\rm lep}_T>15$, this is not enough to extract the signal. The variable $R_{\rm ISR}$ peaks at one for the signal with good enough resolution to reject the background for $R_{\rm ISR}<0.6$ while retaining most of the signal events. Compared to the background, the di-tau invariant mass, $m_{\tau\tau}$, of the signal has a larger slope on either sides of the peak and can reject the $Z/\gamma^*~+$ jets background especially for negative values of this variable. The opening angle between the invisible system and ISR is also effective as the SM background distribution is almost featureless whereas the signal peaks for values greater than 3 rad.

\begin{figure}[H]
\centering
\subfloat[]{\includegraphics[width=0.45\textwidth]{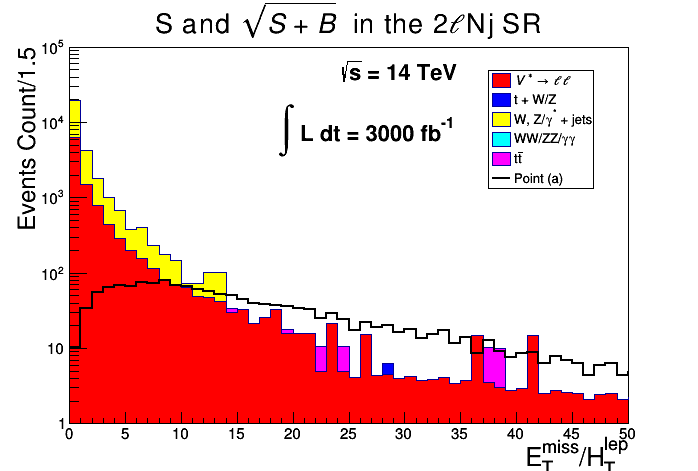}}
\subfloat[]{\includegraphics[width=0.45\textwidth]{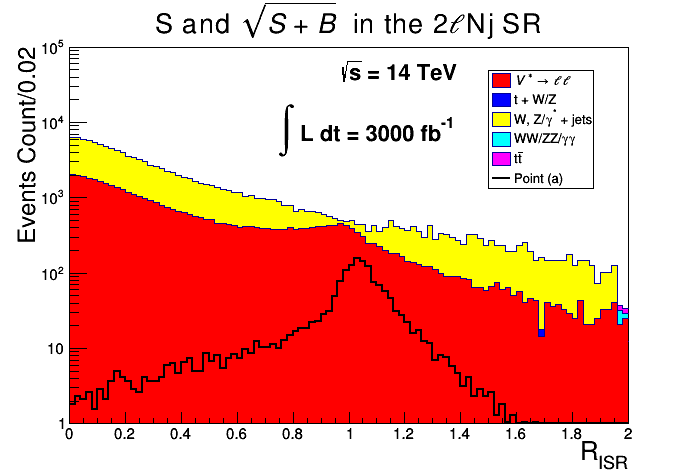}}\\
\subfloat[]{\includegraphics[width=0.45\textwidth]{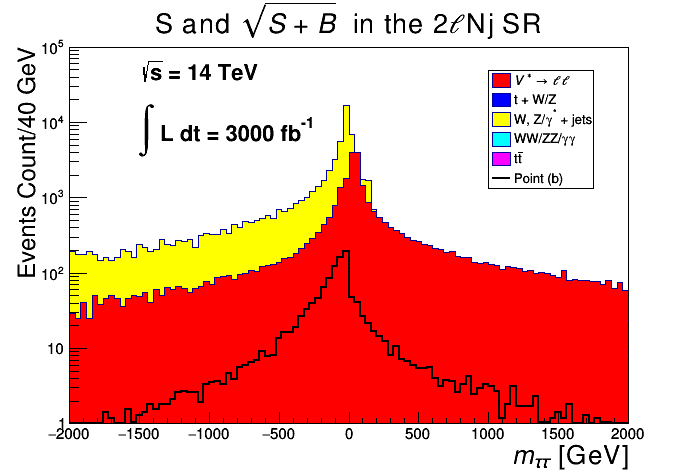}}
\subfloat[]{\includegraphics[width=0.45\textwidth]{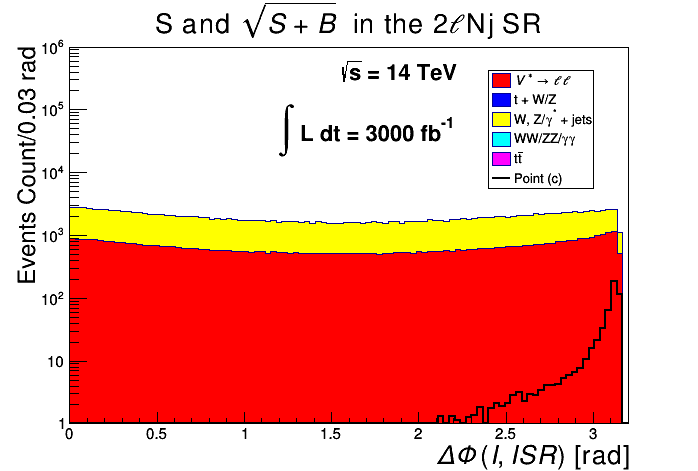}}
\caption{Distributions of four kinematic variables used in this analysis:  (i) $E^{\rm miss}_T/H^{\rm lep}_T$; (ii)  $R_{\rm ISR}$;
(iii) $m_{\tau\tau}$  and (iv) $\Delta\phi(\rm I,ISR)$ for select benchmarks at 14 TeV and for 3000$\ifb$ of integrated luminosity.}
\label{fig5}
\end{figure}

To design effective cuts we look at the two-dimensional distributions in two observables, $R_{\rm ISR}$ and $m_{\tau\tau}$, shown in Fig.~\ref{fig6}. It is clear that for the signal (left panel) most events are clustered near $R_{\rm ISR}=1$ and almost symmetric in $m_{\tau\tau}$ whereas for the background (right panel), events are clustered for smaller $R_{\rm ISR}$ and more negative $m_{\tau\tau}$. This feature can be used to reject the SM backgrounds and enhance the signal-to-background ratio.

\begin{figure}[H]
 \centering
 \includegraphics[width=0.49\textwidth]{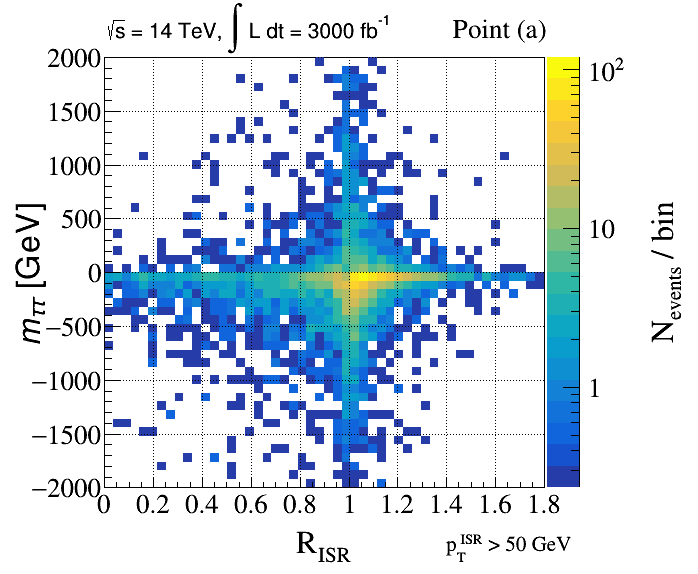}
 \includegraphics[width=0.49\textwidth]{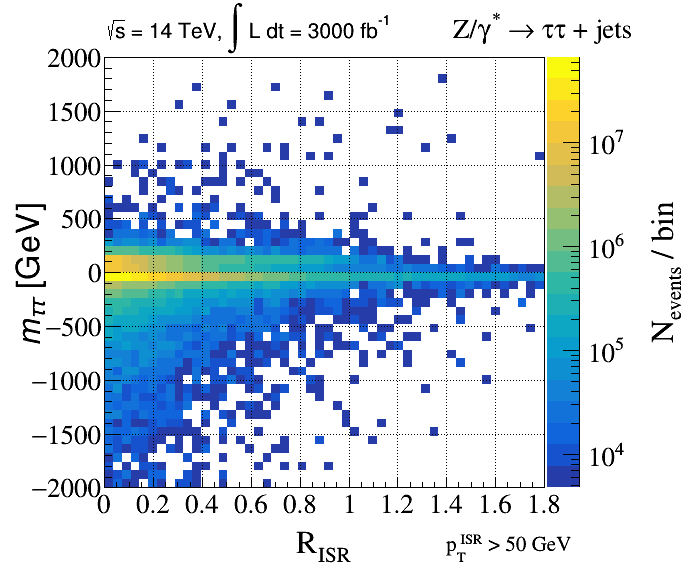}
   \caption{Two-dimensional distributions in $R_{\rm ISR}$ and $m_{\tau\tau}$ for point (a) [left panel] and the dominant $Z/\gamma^*~+$ jets background [right panel].
 Note that for the signal corresponding to model point (a) clustering of events occurs at $R_{\rm ISR}=1$ (left panel) while for
 the background the clustering occurs far away from $R_{\rm ISR}=1$. The simulation is at 14 TeV and for 3000$\ifb$ of integrated luminosity with a 50 GeV cut on the transverse momentum of the leading ISR.}
	\label{fig6}
\end{figure}

We consider two signal regions, SR-$2\ell$Nj-Low and SR-$2\ell$Nj-High targeting low and high mass ranges, respectively. We show the preselection criteria and the analysis cuts used in Table~\ref{tab5}. The preselection criteria and analysis cuts have to be optimized for the 27 TeV case as harder cuts are naturally required to maximize the signal-to-background ratio.

\begin{table}[H]
\begin{center}
\begin{tabular}{l|cc|cc}
\hline\hline\rule{0pt}{3ex}
\multirow{3}{*}{Observable} & SR-$2\ell$Nj-Low & SR-$2\ell$Nj-High & SR-$2\ell$Nj-Low & SR-$2\ell$Nj-High \\
\cline{2-5}
 & \multicolumn{2}{c|}{14 TeV} & \multicolumn{2}{c}{27 TeV} \\
 \cline{2-5}
 & \multicolumn{4}{c}{Preselection criteria} \\
\hline\rule{0pt}{3ex}
$p^{\rm leading~jet}_T$, $p_T^{\ell}$ (GeV) & \multicolumn{2}{c|}{$>30$, $>4$} & \multicolumn{2}{c}{$>30$, $>4$} \\
$E^{\rm miss}_T$ (GeV) & \multicolumn{2}{c|}{$>90$} & \multicolumn{2}{c}{$>100$} \\
$m_{\ell\ell}$ (GeV) & \multicolumn{2}{c|}{$\leq 20$} & \multicolumn{2}{c}{$\leq 40$} \\
\cline{2-5}\rule{0pt}{3ex}
 & \multicolumn{4}{c}{Analysis cuts} \\
\cline{2-5}\rule{0pt}{3ex}
$p^{\rm ISR}_T$ (GeV) & $>50$ & $>50$ & $>80$ & $>80$ \\
$N_j^{\rm ISR}$, $N_j^{\rm V}$ & $\geq 1$ & $\geq 1$ & $\geq 1$ & $\geq 1$ \\
$\Delta\phi(\rm ISR, I)$ (rad) & $>2.8$ & $>2.8$ & $>2.8$ & $>2.8$ \\
$E^{\rm miss}_T/H^{\rm lep}_T$ & $>20$ & $>24$ & $>23$ & $>27$ \\
$R_{\rm ISR}$ & $>0.5$ & $>0.9$ & $>0.5$ & $>0.9$ \\
$m_{\tau\tau}$ (GeV) & \multicolumn{2}{c|}{$>-630$ and $<460$} & \multicolumn{2}{c}{$>-750$ and $<300$} \\
\hline
\end{tabular}
\end{center}
\caption{Preselection and analysis cuts (at 14 TeV and 27 TeV) applied to the signal and SM backgrounds for two signal regions targeting low and high electroweakino mass ranges.}
\label{tab5}
\end{table}

The selection criteria listed in Table~\ref{tab5} are applied to the signal and SM backgrounds simulated at 14 TeV and 27 TeV. The samples are normalized to their respective cross-sections in fb. After the cuts, the surviving signal (S) and background (B) cross-sections are used to determine the integrated luminosity necessary for an $\frac{S}{\sqrt{S+B}}$ excess at the $5\sigma$ level which merits a discovery.  To illustrate the effectiveness of the cuts, we plot the distributions in $R_{\rm ISR}$ for select benchmarks after applying all the cuts in Table~\ref{tab5}, except the ones on $R_{\rm ISR}$ itself. The distributions are shown in Fig.~\ref{fig7}. Panels (i) and (ii) of Fig.~\ref{fig7} show the $R_{\rm ISR}$ distribution for point (a) at 14 TeV (left) and 27 TeV (right). The excess of signal events signifies that point (a) is discoverable at 14 TeV with 500$\ifb$ of integrated luminosity while only 150$\ifb$ is required for discovery at 27 TeV. However, for point (d) shown in panels (iii) and (iv), 500$\ifb$ is not enough for discovery at 14 TeV while this amount is sufficient to claim discovery at 27 TeV.

\begin{figure}[H]
 \centering
\subfloat[]{\includegraphics[width=0.49\textwidth]{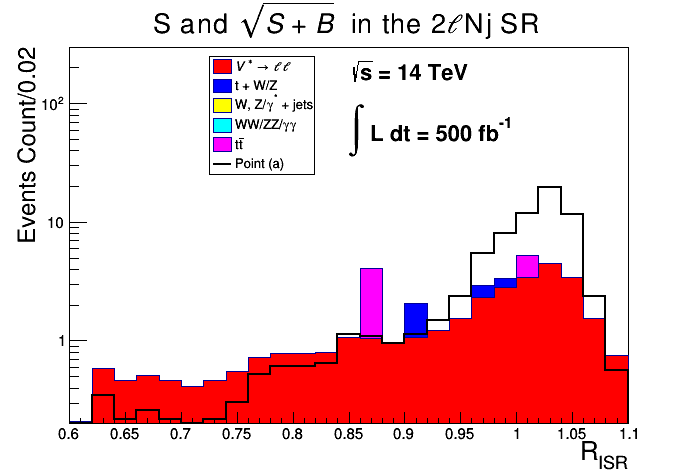}}
\subfloat[]{\includegraphics[width=0.49\textwidth]{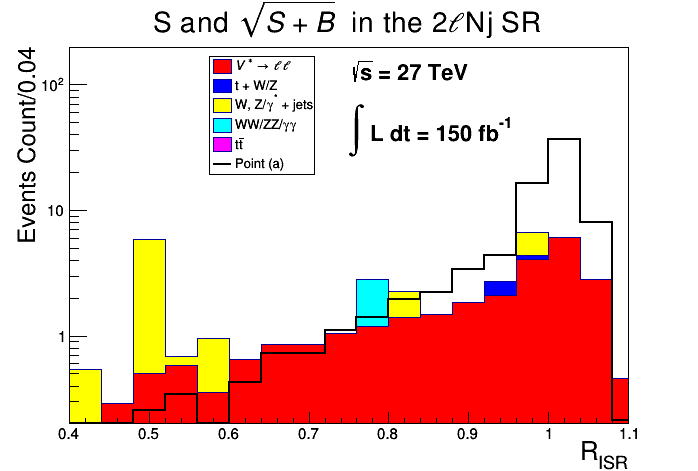}}\\
\subfloat[]{\includegraphics[width=0.49\textwidth]{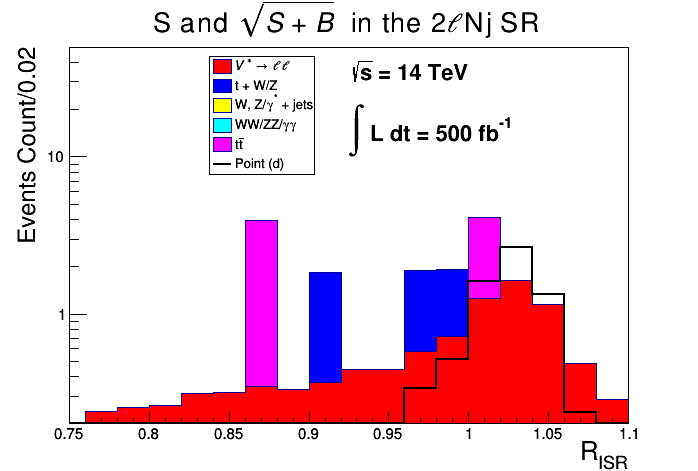}}
\subfloat[]{\includegraphics[width=0.49\textwidth]{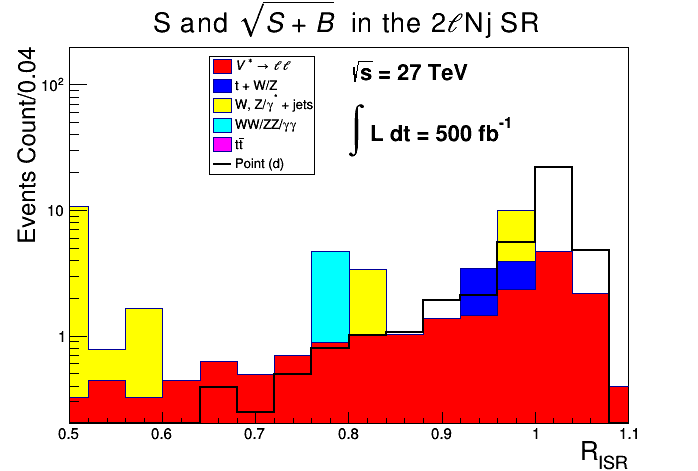}}
 \caption{Distributions in $R_{\rm ISR}$ at 14 TeV and 27 TeV for benchmarks (a) and (d) after applying all cuts in Table~\ref{tab5} except the ones on $R_{\rm ISR}$. Point (a) can be discovered at both HL-LHC and HE-LHC while point (d) is only visible at HE-LHC.}
\label{fig7}
\end{figure}

We calculate the integrated luminosity required for discovery of the ten benchmarks at HL-LHC and HE-LHC. The results are displayed in Fig.~\ref{fig8}. Here one finds that
points (a), (b), (c) and (d) are discoverable at HL-LHC requiring an integrated luminosity of $\sim$ 260$\ifb$ for point (a) and $\sim$ 2060$\ifb$ for point (d). On the other hand, all benchmarks are discoverable at HE-LHC with (a) requiring as little as $\sim$ 70$\ifb$. The integrated luminosities required for this electroweakino mass spectrum range from $\sim$ 70$\ifb$ to 1955$\ifb$ for point (i). Despite having a smaller cross-section, point (j) seems to require slightly less integrated luminosity for discovery compared to point (i). The reason is that the gauginos for this point have a larger mass gap compared to the other points. This advantage allows us to retain more signal events and thus require less integrated luminosity for discovery.

\begin{figure}[H]
 \centering
   \includegraphics[width=0.75\textwidth]{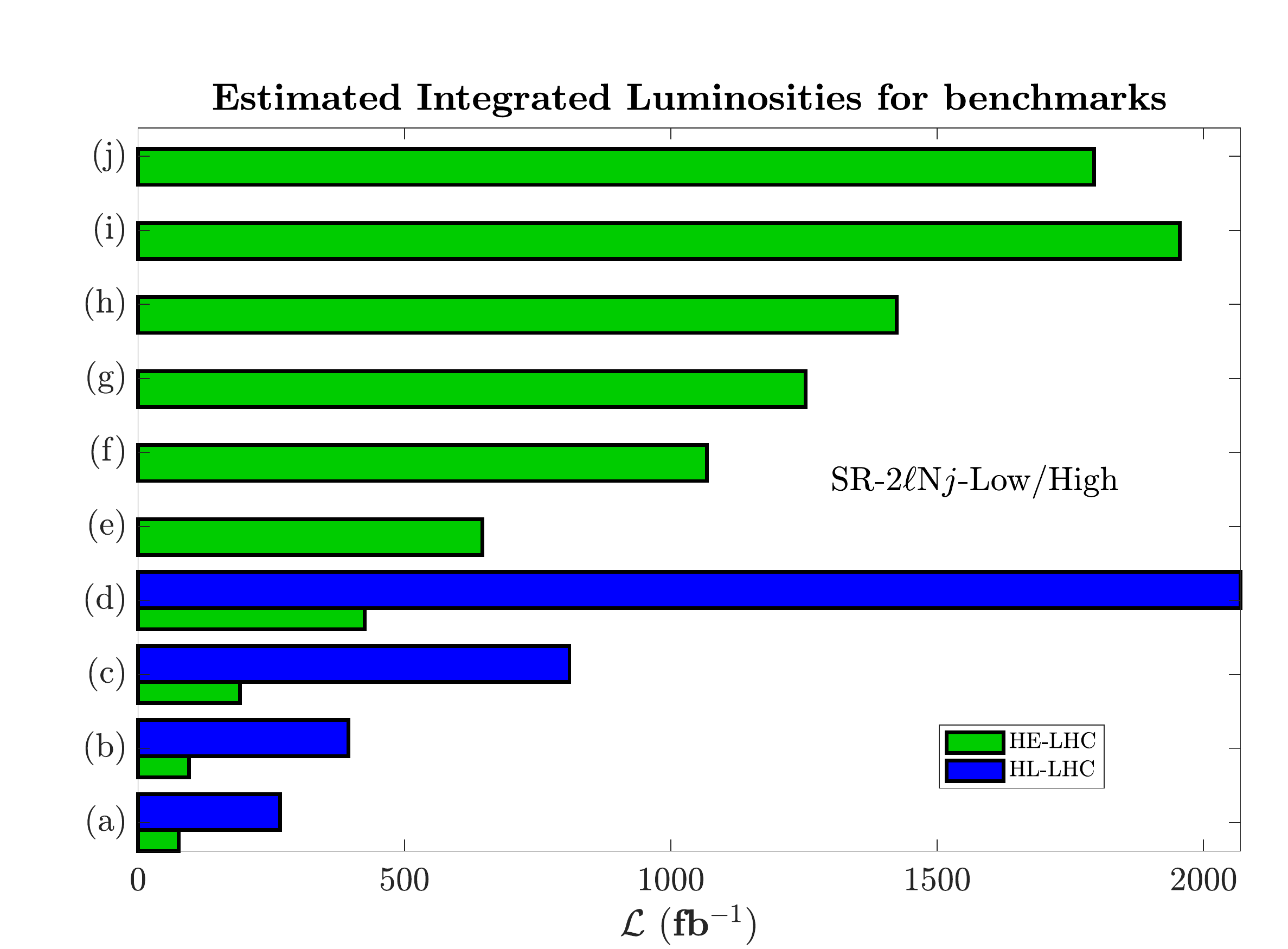}
   \caption{Estimated integrated luminosity for discovery of benchmarks of Table~\ref{tab1} at 14 TeV and 27 TeV. Only four benchmarks are discoverable at HL-LHC while all ten points are visible at HE-LHC.  In the figure SR-2$\ell${Nj}-Low/High are defined as in Table \ref{tab5}.
     }
	\label{fig8}
\end{figure}

An integrated luminosity of 260$\ifb$ should be attainable with run 3 of the LHC, hence point (a) should be discoverable after $\sim 5$ months from resuming operation while points (b)$-$(d) need $\sim 1$ yr to $\sim 6$ yrs. As for the HE-LHC, the rate at which data is expected to be collected is $\sim$ 820$\ifb$/yr which implies that point (a) would be discovered within one month of running, while points (b) to (e) require $\sim 1.5-$9 months and $\sim 1.3-$2.5 yr for the rest of the benchmarks. Thus HE-LHC would be an efficient machine for probing the electroweakino mass range under study.

\section{Conclusion}\label{sec:conc}
The measurement of the Higgs boson mass at $\sim 125$ GeV implies a large size of weak scale supersymmetry lying in the
several-TeV region which makes the observation of supersymmetry at colliders more difficult.  Specifically a large value
of the universal scalar mass in SUGRA models would typically lead to sfermion masses to be large.
However, not all supersymmetric particles need be heavy.
Specifically for models where $\mu$ is relatively small lying in the few-hundred
GeV region, some of the electroweakinos would be light and accessible at the LHC.  Models with small $\mu$ can naturally
arise on the hyperbolic branch of radiative breaking of the electroweak symmetry and thus this branch provides a possible region of the parameter space accessible at colliders.
However, models with small $\mu$ typically imply a significant higgsino content for the
LSP neutralino which leads to copious annihilation of neutralinos in the early universe and consequently
the neutralino relic density significantly below the experimental value.
One possible approach in previous works to correct this problem
is to assume that dark matter is multi-component and use the dark matter candidates
other than the neutralino to make up the deficit.

In this work we propose a solution where the neutralino is the only component of dark
 matter but its relic density arises from more than one source. One source is the conventional freeze-out mechanism
 which, however, produces only a fraction of the desired relic density. To make up the deficit we assume that the
 visible sector couples with a  hidden sector which possesses a $U(1)_X$ gauge invariance and after the
 kinetic mixing and the Stueckelberg mass mixing the neutralinos in the hidden sector mix with the neutralinos in the visible sector by ultraweak
 interactions. While the hidden sector neutralinos are not thermally produced in the early universe,  and we assume that their
 relic density is initially negligible,  they can be produced via interactions of MSSM particles in the early universe.
 For a range of the mixing parameters the hidden sector neutralinos decay into the LSP before the BBN and provide the remaining
 component of the relic density. With the proposed mechanism, models which would otherwise be not viable as they
 do not provide the desired amount of dark matter become viable.

 In this work we have provided a set of benchmarks
 which satisfy the Higgs boson mass constraint, the relic density constraint as well as constraints from the current
 limits on dark matter direct and indirect detection.
 The sparticle spectrum predicted in these models is consistent with the current experimental
 lower bounds. The proposed mechanism enlarges the parameter space of natural supersymmetric models defined by small $\mu$.
 Some of the enlarged
 parameter space of the proposed models may be probed by direct detection experiments while some of the other models
  may be testable at  HL-LHC and HE-LHC. The models considered have  a  very compressed electroweakino spectrum consisting of  charginos and neutralinos which lie in the range 250 GeV to $\sim 870$ GeV.
   However, we show that with appropriate procedures  to suppress the background, some of the parameter points
   are discoverable at the HL-LHC with as low as 260$\ifb$ of integrated luminosity. The discoverable mass range is pushed further to reach $\sim 870$ GeV at HE-LHC with a required integrated luminosity ranging from as little as 70$\ifb$ up to $\sim$ 2000$\ifb$.

\textbf{Acknowledgments: }
The analysis presented here was done using the resources of the high-performance Cluster353 at the Advanced Scientific Computing Initiative (ASCI) and the Discovery and Momentum Clusters at Northeastern University. WZF was supported in part by the National Natural Science Foundation of China under Grant No. 11905158 and No. 11935009.
The research of AA and PN was supported in part by the NSF Grant PHY-1913328.\\

{\bf Appendix: Summary of $A+B\to C+\xi$ processes}\label{AppA}\\

In this appendix we summarize $A+B\to C+\xi$ type processes relevant for the production
of the ultraweakly interacting particle $\xi$.
As noted already in the model we discuss in this work
$\xi$ could be one or the other of the two hidden sector neutralinos $\tilde{\xi}^{0}_{1},\tilde{\xi}^{0}_{2}$.
We note in passing that processes where the final state contains two hidden sector neutralinos will be
doubly suppressed and thus these processes are not considered.
Processes of the type $A+B\to C+\xi$
can be divided into two categories such that the initial particles are either R-parity even or R-parity odd
as exhibited in Table~\ref{tab7}.
Here $f$ stands for any of the three generations of quarks and leptons and
$\tilde f$ for any of the three generations of squarks and sleptons;
$\mathcal{H}$ denotes any one of the states $h,H,A$;
${V}$ denotes neutral gauge bosons, which can be $\gamma, Z,Z^{\prime}$.
Our rough counts gives $\mathcal{O}(10^4)$ processes of
type $A+B\to C+\xi$ using initial and final states listed in Table~\ref{tab7}.
For a typical  $A+B\to C+\xi$  process involving one hidden sector neutralino
 we estimate the relic density contribution to be $\lesssim 10^{-7}$ for values of $\delta$ we use.
Thus
the total contribution of $A+B\to C+\xi$ processes listed in Table~\ref{tab7}
to the dark matter relic density is size $\lesssim 10^{-3}$ which is
significantly smaller than
contributions from $A\to B+\xi$ and $A+B\to \xi$ types of processes in Section~\ref{sec:DM}
for the range of parameters we consider.
The above indicates that  $A+B\to C+\xi$ type processes do not play a significant role in our analysis.

\begin{table}[H]
\newcommand{\tabincell}[2]{\begin{tabular}{@{}#1@{}}#2\end{tabular}}
\begin{center}
\begin{tabular}{|c|c|}
\hline
R-parity & Processes\tabularnewline
\hline
\hline
even &
\tabincell{c}{
$f\bar{f}\to\tilde{\chi}^{0}\xi$, $\tilde{f}\tilde{f}^{*}\to\tilde{\chi}^{0}\xi$,
$f\bar{f}^{\prime}\to\tilde{\chi}^{\pm}\xi$, $\tilde{f}\tilde{f}^{\prime}\to\tilde{\chi}^{\pm}\xi$,
$\tilde{f}\tilde{\chi}^{0}\to\tilde{f}\xi$,
$\tilde{f}\tilde{\chi}^{\pm}\to\tilde{f}^{\prime}\xi$,~\\
$\tilde{\chi}^{0}\tilde{\chi}^{0}\to\tilde{\chi}^{0}\xi$,
$\tilde{\chi}^{0}\tilde{\chi}^{\pm}\to\tilde{\chi}^{\pm}\xi$,
$\tilde{\chi}^{+}\tilde{\chi}^{-}\to\tilde{\chi}^{0}\xi$,
$f+\mathcal{H}/{V}\to\tilde{f}\xi$,~~~~~~~~~~~~~~~~~~\\
$f+H^{\pm}/W^{\pm}\to\tilde{f}^{\prime}\xi$,
$\mathcal{H}\mathcal{H}\to\tilde{\chi}^{0}\xi$, $H^{+}H^{-}\to\tilde{\chi}^{0}\xi$,
${V}{V}\to\tilde{\chi}^{0}\xi$,~~~~~~~~~~~~~~~\\
$\mathcal{H}Z\to\tilde{\chi}^{0}\xi$, $H^{\pm}{V}\to\tilde{\chi}^{\pm}\xi$,
$\mathcal{H}+H^{\pm}/W^{\pm}\to\tilde{\chi}^{\pm}\xi$,
$H^{\pm}+W^{\mp}\to\tilde{\chi}^{0}\xi$.~~~~~~
}
\tabularnewline
\hline
odd &
\tabincell{c}{
$f\tilde{f}\to\mathcal{H}/{V}+\xi$, $f\tilde{f}^{\prime}\to H^{\pm}/W^{\pm}+\xi$,
$f\tilde{\chi}^{0}\to f\xi$, $f\tilde{\chi}^{\pm}\to f^{\prime}\xi$,~~~~~~~~~~~~~~~\\
$\tilde{f}+\mathcal{H}/\to f\xi$,
$\tilde{f}+H^{\pm}/W^{\pm}\to f^{\prime}\xi$,~~~~~~~~~~~~~~~~~~~~~~~~~~~~~~~~~~~~~~~~~~~~~~~~ \\
$\tilde{\chi}^{0}+\mathcal{H}/{V}\to\mathcal{H}/{V}+\xi$,
$\tilde{\chi}^{\pm}+\mathcal{H}/{V}\to W^{\pm}/H^{\pm}+\xi$,~~~~~~~~~~~~~~~~~~~~~~~~~~~\\
$\tilde{\chi}^{0}+H^{\pm}/W^{\pm}\to H^{\pm}/W^{\pm}+\xi$,
$\tilde{\chi}^{\pm}+H^{\mp}/W^{\mp}\to\mathcal{H}/{V}+\xi$.~~~~~~~~~~~~~~~~~
}
\tabularnewline
\hline
\end{tabular}
\end{center}
\caption{An exhibition of
$A+B\to C+\xi$ type processes relevant for producing the ultraweakly interacting particle $\xi$.}
\label{tab7}
\end{table}

\end{document}